\documentstyle[preprint,aps,epsfig]{revtex}
\def\ba{\begin{eqnarray}}
\def\ea{\end{eqnarray}}
\def\beq{\begin{equation}}
\def\eeq{\end{equation}}
\newcommand \Pomeron {I\!\!P}
\begin{document}
\title{Diffraction at HERA, Color Opacity and Nuclear Shadowing
in DIS off nuclei}

\author{L.~Frankfurt}
\address{School of Physics and Astronomy, Tel Aviv University,
         69978 Tel Aviv, Israel}

\author{M.~Strikman}
\address{Department of Physics, Pennsylvania
State University, University Park, PA 16802, USA,\\
and Deutsches Elektronen Synchrotron DESY, Germany\thanks{On leave of
absence from PSU.}}

\maketitle
\begin{abstract}
 
  The QCD factorization theorem for diffractive processes in DIS
is used to derive 
formulae for 
the leading twist contribution to the nuclear 
shadowing of parton distributions 
in the low thickness limit  (due to the coherent projectile (photon) 
interactions with two nucleons).
Based on the current analyzes of diffraction at HERA we
find that the average strength of the interactions which
 govern diffraction in the gluon sector at $x \le 10^{-3},
 Q_0=2 GeV$ is $\sim 50$mb. This
is three times larger than in the quark sector and suggests 
that applicability of DGLAP approximation requires
significantly larger $Q_0$ in  the gluon sector.
We use this 
information 
on diffraction
to estimate the higher order shadowing terms due
to the photon interactions with $N\ge 3$ nucleons
which are important for the scattering of heavy nuclei and to
calculate nuclear shadowing and $Q^2$ dependence of
gluon densities. For the heavy nuclei the  amount of the gluon
shadowing: $G_A(x,Q_0^2)/AG_N(x,Q_0^2)_{\left|\right. x\le 10^{-3}}
\sim 0.25-0.4$  is
sensitive to the probability of the small size configurations within
wave function of the gluon "partonometer" at the $Q_0$ scale.
At this scale for $A\sim 200$ the nonperturbative contribution to the
gluon density is reduced by a factor of $4-5$ at $x \le 10^{-3}$
unmasking PQCD physics in the gluon distribution of heavy nuclei.
We point out that the shadowing of this magnitude would strongly
modify the first stage of the heavy ion
collisions at the LHC energies, and also
would lead to  large color opacity effects in $eA$ collisions at
HERA energies. In particular, the leading twist contribution to
the cross section of the coherent $J/\psi$ production off $A\ge 12$
nuclei at $\sqrt s \ge 70$ GeV is strongly reduced as compared to
the naive color transparency expectations.
The Gribov black body limit for $F_{2A}(x,Q^2)$ 
is extended to the case of the 
gluon distributions in nuclei 
and shown to be relevant for the HERA kinematics 
of $eA$ collisions.  Properties of the final states are also 
briefly discussed.

\end{abstract}
\newpage
\section{Introduction}
It has been  realized by Gribov already before the advent of QCD
that there
exists a deep relation between the phenomenon of high-energy
diffraction and the nuclear shadowing phenomenon
\cite{gribov}.
In particular, the nuclear shadowing due to the
interaction of a virtual photon with two nucleons can
be unambiguously calculated in terms of
the $\gamma^*+N \to X +N$ diffractive  cross section
if the coherence length
$l_c=\frac{2 q_o}{Q^2+M^2}$
is much larger than the nucleus radius, $R_A$.
Here $M^2$ is the invariant mass squared of the
quark-gluon system to which a virtual photon is  transformed \cite{gribov}.
In the case of the charged parton structure functions ($F_{2A}(x,Q^2)$)
connection between shadowing and diffraction has been
 explored for a long time, see
\cite{FS88,kwi,FS89,FLS,NZ,Piller,Kop,FSAGK,Barone,Orsay} and references
therein. The importance of the
color fluctuations-weakly interacting
configurations in the shadowing phenomenon was first understood
in \cite{FS88} where this effect has been estimated
based on the QCD aligned jet model and included in the calculation of
$F_{2A}(x,Q^2)$.

Additional contributions to
the nuclear structure function are  related to the
piece of the photon wave function for which the
coherence length $l_c$ is of the order of the
 average internucleon distance
 $r_{NN}\approx 1.7 Fm$.
These important nuclear effects have been estimated
and explored in \cite{FS88,FLS,Eskola93} using constraints
which follow from the QCD momentum and baryon sum rules.
Account for these effects  leads to a more complicated
QCD evolution which mixes shadowing region and the
region of larger $x$.

In the recent paper \cite{FS981} we started analysis of
the implications of the information which is now available
from HERA on the role of the gluon degrees of freedom in
the diffractive  processes in DIS
for the gluon nuclear shadowing.
We were able to study shadowing for
$x \le 10^{-3}$ and $Q^2 \sim 20 -50 $GeV$^2$
and  predict a factor $\sim 2-3$ larger shadowing for the
gluon channel than for the quark channel. This is in
line with expectations of \cite{FLS93}, though it differs from the
pattern assumed in a number of the models, see e.g. for the recent
summary \cite{eskola}.

 In this paper
we will extend this analysis to  a broad range of $x$ and $Q^2$.
The main tool we will use is
the QCD factorization theorem for the hard diffractive
scattering \cite{Collins}, see also \cite{AFS}.
Application of the QCD factorization
theorem makes it possible 
to establish correspondence between the deuteron(nucleus) infinite momentum 
frame (IMF) and the rest frame descriptions and therefore
to explore advantages of both 
descriptions.

An evident advantage of the IMF description is
the  simple interpretation of the
momentum and the baryon sum rules.
 On the other hand space-time development
of high energy processes and nuclear shadowing phenomenon have
a more clear interpretation within  the nucleus rest frame approach.
Using the QCD factorization theorem \cite{Collins} and the Gribov
analysis of nuclear shadowing we will
derive the model independent expressions for
the leading twist nuclear shadowing of parton densities in the
case of coherent interactions with two nucleons (section 2).
\footnote{For an early discussion of the general arguments for the
presence of the nuclear shadowing in the leading twist and references
 see \cite{FS88}.}
However the shadowing due to the interaction with two nucleons
 cannot diminish the parton density by more than a factor of 0.75
without introducing ghosts into the theory (see discussion in the
end of section 3). Hence
in section 3  we  use the recent analysis \cite{ACW} of  the HERA
diffractive data to extract the  information on the
S-channel dynamics of diffraction which is necessary 
to calculate the effects of coherent  interactions with $N\ge 3$
nucleons. We find out that this analysis implies
 that the  average strength of the interaction responsible for the
diffraction in the gluon channel at the resolution scale
$Q_0\sim 2$GeV and $x \le 10^{-3}$ is
 very large: $\sigma_{eff}\sim 50-60 mb$.
Large value of the interaction strength could be related to the large
cross section of the small color octet  dipole interaction with a
nucleon which is given by
\begin{equation}
\sigma^{inel}_{``color~octet dipole'',N}(E_{inc})={3\pi^2\over
 4}b^2\alpha_s(Q^2)xG_N(x,Q^2\equiv{\lambda\over b^2}),
\end{equation}
where $x={Q^2\over 2m_NE_{inc}}$.  This is a factor of $9/4$ larger than 
for the case of  ``color triplet dipole'' \cite{AFS}.
If we take for $\lambda$ the value we estimated before for the 
color triplet case:
 $\lambda(x \approx 10^{-3})\approx  9$, we find that
the cross section is close to the $S$-channel unitarity limit
for the range of applicability of DGLAP approximation
corresponding to $\sigma_{inel} \ge \sigma_{el}$ for $x\sim 10^{-4}$
and $Q^2 \sim 10$ GeV$^2$. 

For the scattering of a system with a radius $r\ll r_N$ the slope of the 
elastic scattering is given by the square of the
two-gluon form factor $F^2_{2g}(t) \approx \exp(B_{2g}t)$ with 
$B_{2g}\sim 4 GeV^{-2}$.
For this situation condition $\sigma_{inel}=\sigma_{el}$ corresponds to
the  effective cross section of $8\pi B_{2g}=40 mb$. This value
is close to the one 
which emerges from the analysis of the diffractive data where
the size of diffractive system is smaller though not negligible as 
compared to the nucleon size.
Note also that the smallness of the shrinkage of the diffractive cone
for the $J/\psi$ elastic photoproduction 
($\Delta B \le 1 GeV^{-2}$ for $\sqrt{s}$ between 5 and 200 GeV
as compared to $\Delta B \approx 3 GeV^{-2}$ expected in 
the soft regime) indicates that 
perturbative physics occupies most of the rapidity range
for $Q^2\ge 4 GeV^2$ for HERA energy range.

It is worth emphasizing that inapplicability of the DGLAP evolution 
equation  and possible closeness to the unitarity limit 
we discuss here are due to the growth of $xG_N(x,Q^2) $ generated 
predominantly by the $\log Q^2$ terms in the DGLAP evolution equations 
rather than  solely
by the  $\ln (1/x) $ terms which would be the BFKL approximation. 
Hence the pattern discussed here is qualitatively 
different from the BFKL scenario of high-energy dynamics. Indeed,
large values of $xG_N(x,Q^2)$ ($\sim 10-20$ in the small $x$
HERA kinematics and growing with $Q$ as $\sim \sqrt{Q}$)
emerge not because of long ladders in rapidity - the ladders contain 
no more than 2-3 gluons in the multi-Regge kinematics, but rather
due to  a large number of emitters at the lower resolution scale.
Possible closeness to the unitarity limit makes it likely that for moderate 
$Q^2 \le 10 GeV^2$ corrections to the DGLAP 
predictions for the nuclear shadowing would be rather
large. This would primarily affect our predictions for 
moderate $Q^2$ since the information about the gluon induced diffraction 
is obtained predominantly at 
larger $Q^2$ and extrapolated to lower $Q^2$ via the DGLAP equations.

In section 4 we first analyze the dynamics of the fluctuations
of the interaction strength (color coherence - color opacity and 
color transparency phenomena) and explain that
significant  fluctuations of the strength of interaction 
should be present in particular due to the QCD evolution.
Next we  study the nuclear shadowing
originating from the interactions with $N\ge 3$ nucleons. We point out
that the eikonal type  approximation seems
reasonable in the soft QCD regime when
the  projectile wave function contains a large number of
constituents. On the contrary in the PQCD regime where photon
wave function is given by a $q \bar q$ dipole
not more than two inelastic collisions are allowed by energy conservation
law. Otherwise the energy released in the inelastic collisions
calculated through the cuts of exchanged parton
ladders will be larger than the sum of the energies of
the  colliding particles. Evaluation of a larger number of
rescatterings in PQCD is beyond the scope of 
the  naive semiclassical approximation and requires an
accurate account of the space-time evolution of
the  scattering process, in particular
a calculation of the NLO approximation to the photon wave function.
We demonstrate that  the $N\ge 3$ interactions are
sensitive to the  existence of the fluctuations of the interaction strength.
This sensitivity is rather small for $A\sim 12$.
For such $A$ we predict significantly larger shadowing for gluons:
$G_A(x,Q^2_0)/AG_N(x,Q^2_0)_{\left |x \le 10^{-3}, Q_0^2=4 GeV^2\right .}
\sim 0.7$ than for quarks:
$F_{2A}(x,Q^2_0)/AF_{2N}(x,Q^2_0)_{\left |x \le 10^{-3},
 Q_0^2=4 GeV^2\right.}
\sim 0.85$.
For larger $A$ sensitivity to fluctuations steadily increases.
However we find that the average interaction strength  in the gluon
channel is large at the normalization scale of $Q_0=2$ GeV so a
significant nuclear shadowing
of average and larger than average
interaction strengths into the cross section is
determined by the geometry of collisions
and rather
insensitive to the structure of the distribution over the strengths.
As a result of shadowing of
strongly interacting (nonperturbative ?) configurations, 
the relative contribution of the
interactions with small $\sigma$ is strongly enhanced in
the parton distributions in heavy nuclei.
We estimate possible effects of the weakly interacting configurations
and find that they may contribute up to 50 \% to
$G_{A \sim 200}(x \le 10^{-3},Q_0 \sim 2 GeV)$.
\footnote{We are indebted to A.Mueller who stressed the effect of
filtering of the PQCD physics in 
the
parton distributions in nuclei.}
At the same time
the fraction of the cross section due to weakly interacting
configurations should
diminish
with decrease of $x$.

We want to stress here that the use of information on the
diffraction in DIS at HERA allows us to take into
account the nonperturbative effects in the gluon nuclear parton
densities at the boundary of the QCD evolution.
In the previous studies the gluon shadowing either
was treated purely perturbatively as for example
 in the  IMF model of McLerran and Venugopalan \cite{MV} or it was
introduced in a phenomenological way
assuming similarity between the shadowing in the gluon and
quark channels, see e.g. \cite{FLS,eskola}.
Overall a currently popular  scenario which is
used in the discussion of the heavy ion collisions assumes that
reduction of gluon densities is a relatively small correction,
for the recent review and references
see \cite{Mueller99}.

Next, we introduce the constraints
on the gluon densities which follow from the
momentum sum rule and imply presence of the
gluon enhancement at $x \sim 0.1$.
Combing this effect with the quark and gluon shadowing for small $x$
we calculate the $x,Q^2$ dependence of the  leading twist
nuclear densities.
In the end of the section we also consider nuclear structure functions
in the limit when the nucleus thickness is large enough so the
black disk approximation is applicable.

Obviously, the predicted large gluon shadowing has
many implications for the various high-energy processes of
scattering off nuclei. In section 5 we
calculate the impact parameter dependence of the gluon shadowing and
briefly analyze two phenomena: the
emergence of the color opacity in the coherent production of
$J/\psi$ and $\Upsilon$-mesons from nuclei in the HERA kinematics,
and the suppression on the
minijet production in $AA$ collisions at the LHC energies.
We find both the color opacity effect and minijet suppression to
be very large. For example, for the lead-lead collisions
we predict a suppression of the minijet production at
$p_t=2(3)$ GeV/c by a factor $\ge 7 (\ge 4)$.

In section 6 we briefly discuss properties of 
final  states and predict a dip in the ratio of
the spectra of leading hadrons produced in the current fragmentation
region in eA and in eN collisions
for rapidities shifted from the maximum rapidity by 
$\ln \left[ \left<M^2_{diff}\right>/\mu^2\right]$
where $\left<M^2_{diff}\right>$ is the average diffractive mass$^2$
 produced in eN scattering.

In section 7 we compare our approach with some of the recent 
studies of the nuclear shadowing.

\section{The QCD factorization theorem and the
leading twist shadowing for the parton densities.}

The studies of the diffraction production in hard processes lead
 to the introduction of the diffractive parton densities
$f^{D}_{j/B}(\beta,Q^2,x_{\Pomeron},t)$
with  $\beta={x\over x_{\Pomeron}}$,
 which  represent the number densities of
partons in the initial hadron, but conditional on the detection
of the diffracted outgoing hadron $B$ in the target fragmentation
region with light-cone
fraction $1-x_{\Pomeron }$ and fixed momentum transfer $t$.
For example in the case of the diffractive process $e + p \to e + p +X$
the diffractive structure function $F_2^D$ which is introduced
via
\beq
{d^4\sigma_{diff}\over d\beta dQ^2d x_{\Pomeron}dt}=
{2\pi \alpha^2\over \beta Q^4}
\left(\left[1+(1-y)^2\right]F_2^{D}-y^2 F_L^{D}\right),
\eeq
can be written as
\beq
F_2^{D}(\beta,Q^2,x_{\Pomeron},t)=\sum_a e^2_a
\beta f_{a/p}(\beta,Q^2,x_{\Pomeron},t) +HT~ corrections.
\eeq
In the case of the proton production this  structure of the
hard diffractive processes was first
suggested in the framework of the Ingelman-Schlein model \cite{IS}.
Recently it was demonstrated\cite{Collins} that the QCD factorization
 theorem is valid for the  $x,Q^2$ evolution of these parton
 densities at fixed $x_{\Pomeron },t$.
The evolution is governed by the same DGLAP
equations as for the inclusive
processes. The HERA data on diffraction in DIS are consistent with
the dominance of the leading twist contribution except near the edge
of the phase space (see discussion below).

For the processes dominated by the vacuum channel
the Gribov theory \cite{gribov} unambiguously relates diffractive processes in
the scattering of a projectile off a single nucleon to
the process of nuclear shadowing due to the interaction of the projectile
with two nucleons. The simplest way to visualize
this connection for example
 in the case of the scattering off the deuteron is to
consider $\gamma^*d$ scattering in the deuteron rest frame in the
kinematics where
$l_c \gg R_d$ ($R_d$ is the radius of the  deuteron).
Due to the difference of  the spatial
scales characterizing the deuteron and the soft
QCD strong interactions,
the  dominant contribution is given by the diagrams
where the  photon dissociates into
a  hadron component before deuteron and
then this component interacts with both nucleons.
Let us use the AGK theorem \cite{AGK} and
consider the cut of the double scattering diagram corresponding to the
diffractive final state (Fig.1).
This corresponds to the scattering off one nucleon in
the $\left|in\right>$ state and off the second nucleon in the
$\left<out\right|$ state.
The final state interaction  between nucleons
is accounted for as usual
within the closure approximation.
The interference between two diagrams results from the Fermi motion
of nucleons in the deuteron since the spectator nucleon in the
$\left|in\right>$ state has to have a momentum equal to the
momentum of the diffracted nucleon in the $\left<out\right|$ state.
The screening effect is expressed ultimately through
$-Re f^2$ where $f$ is the diffractive amplitude
of the interaction of the probe with the nucleon as
compared to $\left|f\right|^2$ in the case of diffractive
scattering off the nucleon. The real part of the
diffractive amplitude is rather small
and can be calculated from the information
about the imaginary part of the
amplitude. Thus the  difference
between $\left|f\right|^2$ and $-Re f^2$
is small and easy to deal with.

Hence we can apply the Gribov  results for the scattering off the 
deuteron and nuclei to evaluate the shadowing contribution to
the deuteron parton density of flavor $j$ in terms of
the corresponding
nucleon diffractive densities (we consider only the Pomeron type
contribution, so we do not distinguish
diffraction of protons and neutrons)
\begin{equation}
f_{j/^2H}(x,Q^2)=f_{j/p}(x,Q^2)+f_{j/n}(x,Q^2)-
\eta\frac{1}{4\pi}\int dx_{\Pomeron}dt S(4t)
f^{D}_{j/N}
\left(\beta, Q^2,x_{\Pomeron},t\right).
\label{deuteq}
\end{equation}
Here $S(t)$ is the electromagnetic form
factor of the deuteron, and
 $-t=(k_t^2+(x_{\Pomeron}m_N)^2)/(1-x_{\Pomeron})$, and
$\eta=(1-(Re A_{dif}/Im A_{dif})^2)/1+(Re A_{dif}/Im A_{dif})^2)$.

Similarly, in the approximation when only scattering off two nucleons
in the nucleus is taken into account one can similarly deduce
the expression for the shadowing term in terms of the parton densities
\ba
f_{j/A}(x,Q^2)/A  =  f_{j/N}(x,Q^2) -{1 \over 2}
\int d^2b\int_{-\infty}^{\infty}dz_1\int_{z_1}^{\infty} dz_2
\int_x^{x_0} dx_{\Pomeron}\cdot
  \nonumber \\
\cdot f^{D}_{j/N}
\left(\beta, Q^2,x_{\Pomeron},t\right)_{\left|k_t^2=0\right.}
\rho_A(b,z_1)\rho_A(b,z_2)\cos(x_{\Pomeron}m_N(z_1-z_2)).
\label{aeq}
\ea
Here $\rho_A(r)$
is the nucleon density in the nucleus normalized according to
the equation $\int \rho_A(r)d^3r=A$. For simplicity we gave
the  expression
for the limit when the
slope of the
dependence of diffractive amplitude on
the momentum transferred to target
nucleon, $ t$,
is much smaller than the one due to the nucleus
form factor so that impact parameters of two nucleons are equal.
Note that eq.\ref{aeq} is similar to the
corresponding expression for
 the shadowing in the vector dominance model, see eq.(5.4) in
\cite{VDM} since  the space-time evolution
 of the interaction is the same in both cases.
This leads to the same structure of the  nuclear
block, provided one substitutes the VDM expression for
the longitudinal momentum transfer,
$q_z=M_V^2/2\nu$ by the Bjorken limit value:
 $q_z=x_{\Pomeron}m_N$.

The crucial feature of eqs.\ref{deuteq},\ref{aeq} is that the parton
densities which enter in the shadowing term evolve according
to the leading twist evolution equations.
When they are folded with a function
of $x_{\Pomeron}$ which does not depend on $Q^2$ they retain this property.
Since the QCD evolution of real and imaginary parts of hard amplitude
is governed by the same evolution equation
at sufficiently small $x$ we investigate in the paper
the fact that real parts enter into diffraction and into shadowing in
a different way does not influence the QCD evolution.
This proves that eqs.\ref{deuteq},\ref{aeq}
correspond to the leading twist contribution
to the nuclear parton densities. 
{\it In the limit of the low
nuclear densities 
eqs.\ref{deuteq},\ref{aeq}
 provide a complete description of the
leading twist nuclear shadowing.}

Obviously the derived equations could not provide a complete picture
of the deviations of  nuclear parton densities from the sum of the
nucleon densities for all $x$. This is because the derived equations
take into account  the contributions related
to the distances $l_c\gg R_A$ but not the ones related
to the configurations with much smaller coherence lengths.
 The simplest way to estimate the corresponding
additional piece is to apply the energy-momentum
and baryon sum rules which are exact in QCD for the
leading twist parton densities. Therefore
to satisfy these sum rules the shadowing should be
accompanied by an enhancement of some parton densities at higher $x$.
This enhancement
term  has to be added to eqs.
\ref{deuteq},\ref{aeq}. If we introduce this term at a scale $Q_0^2$
for $x\ge x_0$ it would contribute for large $Q^2$  for much smaller
$x$.  Hence the Gribov type approximation becomes inapplicable for
fixed $x$ and $Q^2\to \infty$ \footnote{ In principle one should also
take into account the effects of nonnucleonic degrees of freedom (the
large $x$ EMC effect) but for any practical purposes this effect is
negligible.}.
Below, to deal with the enhancement effects we  will
adopt the procedure of \cite{FLS} in which these effects are estimated
at a low normalization point and  the subsequent evolution is dealt
with by solving the DGLAP evolution equations.

The range of the validity of approximation where interactions with
$N\ge 3$ nucleons are neglected strongly depends on the strength of
the corresponding diffraction channel. Hence in the next two
sections we
review the results of the recent analysis of the HERA diffractive
data and build approximation for treating interactions with several
nucleons.

\section{Diffraction at HERA and shadowing in the low nuclear
thickness  limit}

\subsection{Gap probability for the  gluon induced hard diffraction}
First, let us briefly summarize the results of the studies of
diffraction in DIS which were performed over the last few years
at electron-proton and proton-antiproton colliders and
recast them in
the form necessary for the studies of the nuclear shadowing phenomena.
The data obtained at HERA include studies of the diffractive
structure functions in $\gamma^*+p \to X+p$ scattering,
production of dijets and charm in $\gamma^*+p $ scattering.
The leading twist contribution appears to describe the data well
except very close to the edge of the phase space where higher twist
effects are important. So the factorization theorem for these
processes \cite{Collins} seems to hold for the studied $Q^2$ range,
see \cite{ACW} for the recent analysis.

In the practical applications an assumption is usually made that
the semiinclusive parton densities at small values of
$x_{\Pomeron}$ can be written as a product of
a function of $x_{\Pomeron}$  and
 a parton density which depends on
$\beta=x/x_{\Pomeron}$ and $Q^2$.

 Hence for the sake of brevity we will refer to these densities
as the parton densities in the "Pomeron".
\footnote{Note that in difference from the usual parton densities
which are process independent the 
"Pomeron"
parton densities may
depend on the target, on the mass of diffractively produced system
etc. In particular, for small masses $M^2\ll Q^2$ contributions of
the higher twist to diffraction become important which are proportional to
$xG_N(x,Q^2)^2$ and hence lead to intercept of the effective
''Pomeron'' $\alpha_{\Pomeron}(0)\ge 1.20$ \cite{FMS93}.  In 
the analysis of \cite{ACW}
this kinematics was excluded from the fit. At the same time  for $M^2\gg Q^2$
intercept should be more close to $\alpha_{\Pomeron}(0)=1.08$
familiar from the soft QCD
interactions. These are particular illustrations of the deep
difference between the QCD factorization theorem and the Regge pole
factorization \cite{FKSpi}.
Note also that the energy dependence of diffraction is different from that
for soft hadronic processes. This is not surprising
since the coherence length for the soft hadronic processes is
significantly larger than 
that
for soft hadronic processes in DIS.}
As a result one can define a diffractive parton density at given
$x$ as a convolution of the so called Pomeron flux factor,
$f_{p/\Pomeron}(x_{\Pomeron})$ and corresponding
Pomeron parton density for example for gluons:
\begin{equation}
xg_{dif}(x,Q^2)=\int^{x_{max}}_x
f_{p/\Pomeron}(x_{\Pomeron})x/x_{\Pomeron}f_{g/\Pomeron}
(x/x_{\Pomeron},Q^2)d x_{\Pomeron},
\end{equation}
where $x_{max}$ is the maximal value of $x_{\Pomeron}$ for which
diffractive picture still holds.

The important finding of the HERA diffractive
studies is  that $f_{g/\Pomeron}(\beta)\gg f_{q/\Pomeron}(\beta)$
for a wide range of $\beta$ (Similar trend
 was observed in $p\bar p$ collisions,
see review in  \cite{Dino}).  For example, in the best
global fit of the HERA diffractive data \cite{ACW}
(fit {\bf D}):
\begin{equation}
\beta f_{g/\Pomeron}(x,Q_0^2)=(9.7 \pm 1.7) \beta (1-\beta),
{\Sigma_q\beta f_{q/\Pomeron}(x,Q_0^2)\over
\beta f_{g/\Pomeron}(x,Q_0^2)} \approx 0.13.
\label{fitd}
\end{equation}

Let us consider probability of diffractive events
where the proton remains intact for  the  hard leading twist
processes coupled solely to the gluons. It can be defined as
\begin{equation}
P^g_{dif}(x,Q^2)
= {xg_{dif}(x,Q^2)\over xg_N(x,Q^2)}.
\end{equation}
Since this definition includes only the leading twist contribution
into diffraction it effectively excludes
the contribution of small masses to
the diffraction which could originate from
the higher
twist effects.
Since at small
$x \sim 10^{-3}$ and $Q^2 \sim$ few GeV$^2$ the ratio of
the quark and gluon densities
in a nucleon
is $\sim {1 \over 2}$,  and
$f_{g/\Pomeron}(\beta)\gg f_{q/\Pomeron}(\beta)$
one obviously expects
\begin{equation}
P^g_{dif}(x,Q^2) ={g_{dif}\over q_{dif}}~~
{q(x,Q_0^2)\over g(x,Q_0^2)}~~P^q_{dif}(x,Q^2)
\gg P^q_{dif}(x,Q^2).
\end{equation}

We can quantitatively
estimate $P^g_{dif}(x,Q^2)$ using the fit {\bf D} of \cite{ACW}.
The analysis of \cite{ACW} chooses the initial conditions for the
DGLAP evolution at $Q_0^2=4 GeV^2$ - see eq.\ref{fitd}.
We also take $x_{max}=0.02$ which is
the highest $x_{\Pomeron}$ for which
the one has enough sensitivity to the gluon density in the
"Pomeron". 
However for $x \ll  x_{max}$ the diffraction
probability practically does not depend on the choice of $x_{max}$.
We find that
\begin{equation}
P^g_{dif}(10^{-4}\le x \le 3\cdot 10^{-3},Q_0^2)
\approx 0.34 \cdot (1\pm 0.15),
\label{prob}
\end{equation}
which is much  larger than $P^q_{gap}(x,Q_0^2) \sim 0.12$.
Total probability of rapidity gap which includes double diffractive
events (proton dissociation) is larger by a factor $\sim 1.4$.
This factor can be estimated
assuming the  Regge factorization
for $t=0$ : ${d\sigma(\gamma^* +p\to X_1 +X_{rec}/dt
\over d\sigma(\gamma^* +p\to X_1 +p)/dt}\left.\right|_{t=0} \sim
0.2$ independent of the diffraction
state $X_1$, and taking into account that the slope of the $t$ dependence
in the double dissociation should be smaller by about a factor
of two due to almost complete disappearance
of the proton form factor in the proton vertex. However the cross
section given by the HERA groups includes a small contribution
of the proton dissociation of about
15\% \cite{ZEUSdiff}. So effectively the  scaling factor is
smaller $\sim 1.25$.

Thus in the gluon channel the ratio of total diffraction
to total cross section reaches the value  $\sim $ 0.4
for $Q=2$ GeV.
Thus we conclude the ratio of single diffraction to total
cross section in the gluon channel is close to that for $pp$ collision
(for the soft hadronic processes analogous quantity
is the ratio of sum of the elastic and the single diffraction
cross sections to the total cross sections).

The QCD evolution leads to a decrease of this probability since at
larger $Q^2$ many small $x$ partons originate from ``ancestors''
at $Q_0^2$ with $x \ge x_{max}$ which cannot produce protons with
small $x_{\Pomeron} \le x_{max}$. So at $Q^2=25 GeV^2$, $P^g_{dif}$
drops to about 0.2.
\footnote{Note that experimental studies of the 
"Pomeron" 
gluon densities are
performed either at large virtualities of $Q \ge 5 GeV$ or via scaling
violation of $f^q_{dif}$ for $Q\ge 2 GeV$. So they cannot directly
measure the  large value of $P^g_{dif}$.}

\subsection{Implications for the S-channel picture of
hard diffraction}
The large probability of diffraction in the gluon channel
comparable to that in soft hadron interactions
indicates that configurations involved have large interaction cross
section (here we effectively switch to the $S$-channel language of
 description of diffraction \cite{FP,GW}).
We can quantify this  by using the optical theorem
${d\sigma_{dif} \over dt}\left.\right|_{t=0}={\sigma^2\over 16\pi}$
to introduce the strength of interaction $\sigma_{eff}$ as
\footnote{Here and below we neglect $\le 5\%$ corrections
due to the real part of the amplitude since other uncertainties in the
input are of the order $15-20\%$.}
$^,$
\footnote{For the sake of simplicity we parameterize the $t$
dependence of diffractive cross section as
$d\sigma^{dif}/dt=d\sigma^{dif}/dt_{\left.\right| t=0 }\exp Bt$.}
\begin{equation}
\sigma_{eff}(x,Q^2)\equiv {16\pi d\sigma^{dif}/dt_{\left.\right| t=0 }
\over \sigma} =
 P^g_{dif}(x,Q^2)  16\pi B
\label{sidat}
\end{equation}
The effective cross section $\sigma_{eff}(x,Q^2)$ characterizes
within the Gribov theory the  diffractive rescatterings of
the produced quark-gluon system,  cf. eq.\ref{shad2}.
The results of the calculation are presented in Fig.2
for $Q=2 $GeV and  $x_{max}=0.02$ for
the quark and gluon channels and show that
$\sigma_{eff}$ for the gluon channel is of the order 55 mb for
small $x$ and about 3 times larger than for the quark channel.
\footnote{Determination of $\sigma_{eff}$ requires the knowledge of
the $t$-dependence of the diffraction. Experimentally it was measured
for the process $\gamma^*+p \to X+p$ only  \cite{slopezeus}
and for relatively large
$\left<x_{\Pomeron}\right> \sim 0.01$. For this kinematics the fit of
\cite{ACW} describes the $t$-dependence well. The fit also assumes the
rate of the diffractive cone shrinkage $\propto \exp(2\alpha'\ln
(1/x_{\Pomeron}))$ with $\alpha'$ from the soft processes.
Due to a larger value of
diffraction and hence a larger value of $\sigma_{eff}$ for the gluon
channel  the assumption of \cite{ACW} that the slope for
single diffraction in
the gluon channel is at least as large as for the quark channel seems
also very natural.  One should however remember that
the lack of direct measurements of the
$t$-dependence of  the gluon induced diffraction introduces an
additional uncertainty in the results of calculations. Overall our
guess for the uncertainty in the value of the parameter $\sigma_{eff}$
for $10^{-4} \le x \le 10^{-3}$ is about 20\%.} Large value of
$\sigma_{eff}$ can be interpreted as
an indication that the interactions in the gluon channel is remaining
strong up to much larger
virtualities than in the quark channel. This matches rather
naturally with the perturbative QCD pattern of a factor of 9/4 larger
cross sections for  color octet dipole -nucleon interaction than for
the color triplet dipole-nucleon interaction
\cite{FLS93},\cite{AFS}, see discussion in the introduction after eq.1.

\subsection{$\sigma_{eff}$ and shadowing for small nuclear densities}

Now we are in a position to rewrite
the results of section 2 for the parton shadowing in the
limit of small nuclear densities using the notion of $\sigma_{eff}$.
This would allow us as a next step to go beyond
the  two nucleon approximation for the shadowing effects.
For  $x\ll x_{max}$, and
not too large $A$ such that $l_c \gg R_A$
the $\cos(x_{\Pomeron}M_N(z_1-z_2))$ factor in eq.\ref{aeq}
can be substituted by one.
Hence in this limit
 the amount of shadowing is directly proportional to
the differential probability of diffraction at $t=0$:
\begin{equation}
xG_A(x,Q^2)=A G_N(x,Q^2)-4\pi\int d^2b T_A^2(b)
\int^{x_{max}}_x {df(x_{\Pomeron})\over dt}\left.\right|_{t=0}
f_g(x/x_{\Pomeron},Q^2)d x_{\Pomeron},
\label{shad2}
\end{equation}
  where
\begin{equation}
T_A(b)=\int_{-\infty}^{\infty}dz \rho_A(z,b),
\label{tb}
\end{equation}
is the usual nuclear thickness function and $\rho_A(r)$ is the nuclear
density. For larger $x_{\Pomeron}$ one has to take into account
the damping factor due to the $\cos(x_{\Pomeron}M_N(z_1-z_2))$ factor
which originates from the nuclear
form factor due to the longitudinal momentum
transfer in the transition to the diffractive mass equal to
$m_Nx_{\Pomeron}$. However with our choice of $x_{max}=0.02 $ this
effect is small even for $A\sim 200$. Eq.\ref{shad2} leads to the
shadowing proportional to $\sigma_{eff}$:
\beq
1-xG_A(x,Q^2)/AxG_N(x,Q^2)={\sigma_{eff}\over 4A}\int d^2b T_A^2(b)
\eeq
and hence predicts a factor $\sim 3$ larger shadowing for the
gluon channel than for the quark channel for $Q=2 GeV$. This is in
line with expectations of \cite{FLS93}, though it differs from the
pattern assumed in a number of the models, see e.g. for the recent
summary \cite{eskola}.

Note also that if one does not separate leading and higher twist
 contributions in the diffraction  off a nucleon one can still use
the experimental data about the total cross section
of diffraction off a nucleon to calculate using
eq. \ref{shad2} the total  amount
of shadowing in the corresponding
 channel for the scattering off the deuteron (and nuclei
in the approximation when the
interactions with $N\ge 3 $ nucleons are neglected).
The simplest way to see this is to apply the
the AGK cutting rules \cite{AGK} which are valid for the scattering
off nuclei. In particular, if the
the higher twist effects due to interactions with two nucleons
described by the Mueller and Qiu model \cite{MQ}
were important in the nuclear shadowing at the normalization point
they should be manifested as well in the diffraction off a nucleon.
So, as far as the diffraction is described by
the leading twist factorization approximation, eq. \ref{shad2}
leads to the DGLAP evolution of the  nuclear  shadowing.

Note also that in the approximation when
a probe (photon) may interact not more than with two nucleons
there exists  a relation
 between the shadowing for the total cross section and
the partial cross section of inelastic processes with the
multiplicity similar to the one in the inelastic $ep$
scattering - $\sigma_1$:
\beq
\sigma_{tot}=\sigma_{imp}-\sigma_{double}, ~~
\sigma_{1}=\sigma_{imp}-4\sigma_{double},
\label{single}
\eeq
where $\sigma_{imp}$ is the impulse approximation cross section and
$\sigma_{double}$ is the screening cross section due to
the interaction with two nucleons \cite{AGK}.
One can see from eq.\ref{single} that
shadowing due to the photon interaction with two  nucleons
can diminish the total
cross section by not more than a factor of 0.75 as compared
to the impulse approximation without introducing ghosts into the
theory: for $A_{eff}/A \le 0.75$
the partial  cross section $\sigma_{1}$
 would become negative \cite{FKS96}.\footnote{Note that in Ref.\cite{EQW}
where shadowing was calculated in the Mueller and Qiu model the values of
$A_{eff}/A$ as low as 0.5 were obtained.}
This implies that for $A_{eff}/A \le 0.75$ shadowing interactions
with a larger number of nucleons could not be ignored.

\section{Fluctuations of the interaction strength and multinucleon shadowing}
\subsection{Modeling effects of cross section fluctuations}

Due to the large value of $\sigma_{eff}(x,Q^2) $
for the gluon channel, deviations
from eq.\ref{shad2} due to interactions with $N\ge 3$ nucleons
become large already for $A\sim 10$.
To account for these effects we address
the $Q^2$ dependence of $\sigma_{eff}(x,Q^2)$.
Within the DGLAP approximation it basically reflects an influx to
small $x$ of  configurations which at a lower resolution $Q^{\prime}$
correspond to configurations with larger $x\equiv x_{parent}$
and hence with smaller $\sigma_{eff}(x_{parent},Q^{\prime})$.
Configurations which interacted strongly at $Q\sim
Q_0$ interact strongly at large $Q$ as well, but they contribute
smaller and smaller  fraction of the total cross section
relevant for the nuclear shadowing phenomenon
at a fixed $x$.
This pattern is the same as in the QCD aligned jet model \cite{FS88}.
Since the gluon shadowing strongly reduces
gluon densities already $Q \sim Q_0$
the deviations from the DGLAP equations for $Q\sim Q_0$
due higher order order terms in nuclear parton density
originating from the average masses of the diffractively produced
system should be significantly
smaller than in the model of \cite{MQ} where shadowing
at the starting scale is neglected.
One can speculate of course that these effects have
already occurred between $Q^2 \sim 1  GeV^2$ and $Q^2=Q_0^2$. Higher
twist effects are enhanced  for the contribution of diffractively
produced system with the masses $M^2\ll Q_0^2$.  This interesting
question is beyond of the scope of this paper.

It is straightforward
to take into account effects
of the longitudinal momentum transfer in the diffraction
\cite{FS89,FS981}. However as we demonstrated in \cite{FS981}
that these effects are important only for $x_{\Pomeron} \ge 0.03$
for $A\sim 200$ and even for larger $x_{\Pomeron} \sim 0.05$
for light nuclei. Since we have chosen $x_{max}=0.02$ we can safely
neglect this effect in the following
discussion. In this approximation
to account for the fluctuation effects it is
convenient to introduce the probability distribution over the
strength of interaction in the gluon channel - $P_g(\sigma)$.
$\sigma_{eff}$ is expressed in terms of $P_g(\sigma)$ as
\cite{FGS}
\begin{equation}
\sigma_{eff}=\int d \sigma \sigma^2P_g(\sigma)/
\int d \sigma \sigma P_g(\sigma).
\label{seff}
\end{equation}
We obtain
in the generalized eikonal approximation:
\begin{equation}
{G_A(x,Q_0^2)\over G_N(x,Q_0^2)}={\int d^2b d \sigma P_g(\sigma)
(2-2 \exp(-T(b)\sigma/2)) \over A \int d\sigma \sigma P_g(\sigma)},
\label{fluct}
\end{equation}
where $T(b)$ was defined in eq.\ref{tb}. Fluctuations lead to a decrease
of shadowing effect as compared to the
quasieikonal approximation where
$P(\sigma) \propto \delta(\sigma-\sigma_{eff})$\footnote{Note that since
the value of $\sigma_{eff}$ is fixed by the cross section of the
diffractive scattering this approximation differs from the eikonal
 approximation often used for the hadron-hadron scattering in
which only elastic rescatterings are included.} We will
study effects of fluctuations  at
length elsewhere. However  for characteristic
$\sigma \sim \sigma_{eff} \sim$ 55 mb
the exponential factor in the numerator of eq.\ref{fluct} is very small for
$A\sim 200$ and small enough $b$ a wide range of $\sigma$, leading to
cross section $\approx 2\pi R_A^2$.  This suppresses
the contribution of large $\sigma$ -nonperturbative QCD physics
and therefore enhances the contribution of small $\sigma$-PQCD
physics.

As a result of large absorption for $\sigma \sim \sigma_{eff}$
 fluctuations near the {\bf average}
value of $\sigma$ practically do not change the shadowing (for
$A\le 250$) provided $\sigma_{eff} $ is kept fixed. If, for example, we
assume that $P(\sigma)= a\theta(\sigma-\sigma_0)$ the $G_A/G_N$ would
change even for heavy nuclei by less than 20\%.

Besides, as we mentioned in the
introduction the  diffraction in the gluon induced hard processes
without proton break up constitutes about 35\% of the events
at $x \le 10^{-3}$.
Large contribution of small $\sigma$ in
$\int d\sigma \sigma P_g(\sigma)$ would imply that
for scattering in a significant fraction of configurations the gap
probability is much smaller than average. Hence some configurations
would have to generate the gap events with a probability
exceeding 35\%.
This is one of indications for the problems for the applicability of
DGLAP to describe $ep$
scattering at HERA at low normalization point $Q_{0}^2$.
At the same time there is a natural mechanism in PQCD for the
generation of a contribution of very small $\sigma$ in
$\int d\sigma \sigma P_g(\sigma)$. It comes from the QCD evolution.
The partons at given $x$ which originated from
ancestors  at lower resolution  scale with $x_{initial} \ge x_{max}$
do no contribute to diffraction and hence effectively correspond to
$\sigma \sim 0$
(In other words this effect reflects the contribution of small coherence
lengths to the small $x$ physics due to
the
QCD evolution).
This effect is especially pronounced in the models of QCD
evolution with low normalization point like the GRV model \cite{GRV}.
To estimate the sensitivity to this effect we will use two models.
The first one
is the  quasieikonal
which neglects higher order fluctuations.
 The second model is
the fluctuation two-component model (to which we will refer
to as a fluctuation model) which implements an extreme
assumption that a fraction
$\lambda$ of the total cross section at given $x$ originates from
configurations with small cross section and the rest from the average
ones. The second model is similar to the QCD aligned jet model of
\cite{FS88,FS89}. It corresponds to
\begin{equation}
\sigma P_g(\sigma)
\propto
\lambda \delta(\sigma) + (1-\lambda)
\delta(\sigma-\sigma_0),
\end{equation}
where $\sigma_0$ is fixed by eq.\ref{seff} to
\begin{equation}
\sigma_0={\sigma_{eff}\over 1-\lambda}.
\end{equation}
The requirement that the gap probability for  this component
(including dissociation of the nucleon) does not
exceed 50\% puts an upper limit on $\lambda$. Taking into
account uncertainties
in the value of the total rapidity gap probability we estimate
that $\lambda \le 0.2$. So we will use $\lambda=0.2$ in this model.
Note that for small values of $\sigma_{eff} \le 20 mb$
and moderate values of $\lambda$ the screening weakly depends on
$\lambda$. Hence we will ignore this effect for the quark
channel.
For simplicity we will also assume that $\lambda $ does not depend on $x$.
This should be considered as a rather rough approximation since for
small enough $x \le 0.005$ one may expect the contribution of
small $l_c$ to decrease with decrease of $x$. Also
for larger $x$ the  drop of the parameter $\sigma_{eff}$ can be due to
an increase of $\lambda$. However in this region we anyway have
significant uncertainties due
to the contribution of $x_{\Pomeron} \sim x_{max}$.
Thus the   results in this $x$ range can be considered
simply as a smooth interpolation between the
region of large shadowing and the region where shadowing disappears.

Obviously,  small $\sigma$'s in eq.\ref{fluct} would
give the dominant contribution for $A\to \infty$. However
for large $\sigma_{eff}$ and $\lambda \le 0.2$
for $A\sim 240$ a contribution of the weakly interacting
component could at most become comparable to the
soft contribution, see Figs.3,4 below.

\subsection{Models for  shadowing at $Q_0^2$ and numerical results
for $Q^2$ dependence}

Based on the above discussion of the fluctuation effects we adopt
the following prescription in the
calculation for the gluon and quark  nuclear shadowing for $A\ge 10$:

(i) For $Q=Q_0 $ we use two models: the quasieikonal model and
the fluctuation model with $\lambda=0.2$
analogous to the ones we used in \cite{FS89,FS981}.
The relation between the amount
of shadowing in the two models is given by
\ba
G_A(x,Q_0)/G_N(x,Q_0)_{fluct.mod.}(\sigma_{eff}(x,Q_0)= \nonumber \\
\lambda + (1-\lambda)G_A(x,Q_0)/G_N(x,Q_0)_{quasieik.mod.}
({\sigma_{eff}(x,Q_0)\over 1-\lambda})
\ea
(ii) To constrain the behavior of the gluon density
at $x \ge 0.02$ in the normalization point
we use the analysis of \cite{FLS} which indicates that gluons in nuclei carry
approximately the same fraction of the momentum as in a free nucleon:
\beq
\int_0^1 dx xG_A(x,Q^2) \approx \int_0^1 dx xG_N(x,Q^2).
\label{sumrule}
\eeq
This allows to estimate the amount of the gluon enhancement at $x\ge
x_{max}$ assuming that it should be concentrated at $x\le 0.2$ where
average longitudinal distances (the Ioffe distances) contributing to  the
parton density are comparable to the internucleon distances -
this procedure is similar to one we introduced in \cite{FLS}.

(iii) Based on the rational presented above we use
the DGLAP evolution equations to calculate the
nuclear shadowing for larger $Q^2$.

Since in this paper we are interested primarily
in the behavior  of the nuclear  gluon and quark densities  at $x\le
0.01$ we are not sensitive to details of the enhancement pattern.
So we
do not try to introduce an $A$ dependent shape for the enhancement
and assume that the enhancement is present for $0.02 \le x \le 0.2$
and can be approximated by
 $G_A(x,Q_0)/G_N(x,Q_0)=C(A)(x-0.02)(0.2-x)$.
We also do not model small enhancement for $F_{2A}(x,Q^2)$ and
$V_A(x,Q^2)$ at $x\sim 0.1$.
In the calculations we use the standard Fermi step fit
to the nuclear densities: $\rho(r)=C/(1+\exp((-R+r)/b))$, where
$R=1.1fm\cdot A^{1/3},b=0.56 fm$.
First, in Fig.3 we present a comparison of the gluon shadowing
calculated in two models at $Q_0=2$GeV and $x=10^{-3},10^{-4}$. One
 can see that shadowing for $A\sim 12$ is already very significant,
though the effect of fluctuations is still small. With a
further increase of $A$ the  effect of fluctuations
becomes larger and it reaches a factor of
1.4  for A$\sim $ 200. For these $A$ in the
fluctuation model weakly interacting configurations contribute
approximately half of the cross section.
Note however that we expect that in a more realistic model
of fluctuations a relative contribution of these
contributions is likely to decrease with decrease of $x$, see
discussion in section IV.A.
  In Figs.4,5 we present
results for the $Q^2$ dependence of
shadowing for gluons and for quarks
for A=12, 40, 100, 200 calculated in the
quasieikonal model.  One can
see that the gluon shadowing is large already for A=12 and for heavy
nuclei reaches the level of $A_{eff}/A \sim 0.2$. (Variations of the
parameter $\sigma_{eff}$ within a factor of $1 \pm 0.2$ allowed by the
uncertainties in the value of the gluon diffractive density at $t=0$
lead to the similar variations of $G_A/AG_N$ at $x \le 10^{-3}$ and of
$\left[1-G_A(x,Q^2)/AG_N(x,Q^2)\right]$ for $0.02 \le x \le 0.2$.)
Shadowing decreases with increase of 
$Q^2$ but  remains large up to very large
$Q^2$.   The shadowing in the charged parton channel is much smaller
and rather weakly decreases with $Q^2$.  In fact, at $x \sim 10^{-4}$
shadowing for $F_{2a}(x,Q^2)$ first increases with increase of $Q^2$
due to a larger shadowing in the gluon channel.   Note also that
eq.\ref{sumrule} leads to a rather large enhancement of $G_A/G_N$ for
$0.05 \le x \le 0.2$.  However the enhancement is
rather large already for
$A=12$. As a result a further growth of the enhancement between carbon
and tin is of the order of 10\% and it is well consistent with the
analysis of the current data in \cite{Pirner}.  The decrease of
shadowing  with increase of $Q^2$ is due to the feeding of the small
$x$ by partons which originated from $x \ge 0.02$ at $Q_0^2$.
Effectively, 
as
we discussed above, the QCD evolution leads to 
fluctuations in the value of interaction strength
due to the mixing of  the contributions of small and large $l_c$.
 Therefore  the use of eq.\ref{fluct} in the
quasieikonal model with $\sigma \propto \sigma_{eff}$ defined at a high
resolution $Q^2$ would lead to an
overestimate of the nuclear shadowing since the cross
section fluctuations lead to a decrease of the shadowing for the fixed value
of $\sigma_{eff}$, see discussion in \cite{FS981}.

Note also that 
it is often stated in the literature that nuclear shadowing for the total
cross section of  $\gamma^* A $ should be practically 
the same for the real photon and for virtual photons with moderate
(few GeV$^2$) virtualities.
However our analysis predicts a significant 
drop of the nuclear shadowing in
the total cross sections of $\gamma^* A $ scattering between
$Q^2=0$ and $Q^2=4 GeV^2$
due to the strong  $Q^2$ dependence of the diffraction contribution 
to $\sigma_{\gamma^*N}$. The results of our calculation using data
on the diffraction in the $\gamma p$ scattering at $W\sim 14 GeV$ 
\cite{Chapin} 
and at HERA \cite{H1} and a smooth interpolation between 
two energy ranges is presented in Fig.6. For a rough estimate we use
the quasieikonal  approximation which leads to a slight 
overestimate of shadowing for the heavy nuclei.

\subsection{Structure functions of nuclei in the the black disk limit}

It is of interest to consider also the structure functions in the
limit $Q^2$=const, $x\ll {1\over m_NR_A}$, $A\to \infty$ first analyzed
by Gribov for $\sigma_T,\sigma_L$ \cite{gribov}.
It was argued in this paper that in such a limit interactions of all
essential configurations in the virtual photon wave function with a
nucleus can be treated in the black body ($S$-channel unitarity) limit. 
Since in the  black disc limit the   dispersion over the strengths of
interactions can be neglected one finds \cite{gribov}
\beq
{1\over Q^2}F_{2A}(x,Q^2)={\pi R_A^2\over 12\pi^2}
\int_{m_o^2}^{W^2\over 2m_N R_A}
{m^2 \rho(m^2) dm^2 \over (m^2+Q^2)^2},
\eeq
and
\beq
{1\over Q^2}F_{L~A}(x,Q^2)={\pi Q^2 R_A^2\over 12\pi^2}
\int_{m_o^2}^{W^2\over 2m_N R_A}
{\rho(m^2) dm^2 \over (m^2+Q^2)^2},
\eeq
where $\rho(m^2)=\sigma(e^+e^-\to hadrons)/\sigma(e^+e^-\to
\mu^+\mu^-)$.

We can generalize
the Gribov formulae deduced for the sea quark distribution \cite{gribov}
to the case of the gluon channel:
\beq
xG_A(x,Q^2)/Q^2=\frac{3}{2}\frac{\pi R_A^2}{12\pi^2}
\int_{m_o^2}^{W^2\over 2m_N R_A} \tilde{\rho}(m^2){m^2dm^2 \over
(m^2+Q^2)^2} \label{blackdisc} \eeq Here
$\tilde{\rho}(m^2)=\sigma(\hat{J} \to hadron)/\sigma(\hat{J} \to 2
g)$ is the ratio of the cross section of the gluon hadronic processes
initiated by the local operator 
$\hat{J}={1\over \sqrt{-\Box}}F_{\mu \lambda}^{\alpha} F^{\mu 
\lambda}_{\alpha}$ 
introduced in \cite{MV} 
to the
perturbative cross section of the annihilation of the gluon source into
gluons.  For $m^2\to \infty $, $\tilde{\rho}(m^2)=1$. So in the black
disk limit
\beq xG_A(x,Q^2)/Q^2= \frac{\pi R_A^2}{8\pi^2} \ln ({W^2\over
Q^2 2m_N R_A})\equiv \frac{\pi R_A^2}{8\pi^2}\ln\left({x_0\over x}\right),
\label{black}
\eeq
 where $x_0={1\over 2 m_NR_A}$.
Eq.\ref{black} provides a solution
for the problem of the gluon nuclear shadowing in the theoretical limit
$Q^2$=const, $x\ll x_0$, $A\to \infty$.
It also  illustrates
 that investigation of the $Q^2$ dependence of the  nuclear
structure functions will provide an effective method to establish
whether fluctuations of strengths play a significant  role
in the gluon  structure
functions.

It is useful to compare eq.\ref{black} with 
the impulse approximation value of $G_A=AG_N$. For example 
for  $A=200$ this equation leads to $xG_A(10^{-4},10GeV^2)/A=8$
 while the current fits to the nucleon data lead to 
$xG_N(10^{-4},10GeV^2)\approx 20$. 
Hence the model independent unitarity constrain implies 
a large gluon shadowing for this kinematics. This is consistent with
our model calculations presented in the  previous subsection.

\section{Onset of color opacity regime in  hard diffraction and
suppression of minijet production in  nucleus-nucleus collisions}

Let us now briefly discuss some of the consequences of the
found magnitude of the gluon shadowing.

The production of minijets is often considered as an effective
mechanism of producing high densities in the head on  heavy ion
collisions. However in the LHC kinematics for the central rapidities
minijets are produced due to collisions of partons with
$x_{jet}={2p_t\over \sqrt{s_{NN}}}$. For heavy ion collisions
$s_{NN} \ge 4~ TeV$ and the gluon-gluon collisions are responsible for
production of most of the  minijets. Therefore the gluon nuclear shadowing
would lead to a reduction of the rate of the jet production
due to the leading twist mechanism by a large
factor up to $p_t\sim 10 GeV/c$, see Fig.7 where we give results
of calculation in the quasieikonal and fluctuation models.
The nuclear gluon shadowing leads to a
similar very strong reduction of
the heavy onium production in $pA$ and $AA$ collisions at LHC
energies for $y_{c.m.}\sim 0$ and small $p_t$.

For the central impact parameters the reduction is even larger.
Using generalized eikonal
 approximation of  eq.\ref{fluct} we can calculate also the
suppression of
the parton densities at a given impact parameter $b$ as
\cite{FLSbnl}:
\begin{equation}
G_A(x,Q_0^2,b)_{shadowed}={\int d \sigma   P_g(\sigma)
(2-2 \exp(-T(b)\sigma/2) G_N(x,Q_0^2)
\over \int P_g(\sigma)\sigma d\sigma}.
\end{equation}
Since in the impulse approximation
$G_A(x,Q^2,b)_{imp}=G_N(x,Q^2)
T_A(b)$ we finally obtain
\begin{equation}
G_A(x,Q_0^2,b)/G_N(x,Q_0^2)={\int d \sigma   P_g(\sigma)
(2-2 \exp(-T(b)\sigma/2) \over T(b) \int d\sigma \sigma
P_g(\sigma)}.
\label{fluct1}
\end{equation}
The results of calculation using eq.\ref{fluct1} are presented in
Fig.8.

So we conclude that the shadowing effects are likely to reduce
very substantially the parton densities generated at the first stage
of heavy ion collisions at  the LHC energies.
However theoretical uncertainties related to 
the role of point-like configurations
lead to rather large uncertainties in the estimate of the suppression
for the case of  the heavy ion collisions. Hence
it would be very important to perform
a direct measurement of the gluon nuclear shadowing at HERA in this
kinematics.

Presence of a large gluon shadowing leads to large effects in the 
diffractive $eA$ collisions at HERA energies.
Here we consider  the simplest example - coherent diffractive production
of vector mesons at large $Q^2$ by the longitudinally polarized photons -
$\gamma^*_L+A \to V +A$,
and photo(electro) production of heavy onium states.
Since these processes are dominated by production of $q\bar q$ in a
small size configuration one may naively expect that
color transparency should hold for such processes and  the amplitudes
of these processes should be proportional to $A$.
However the QCD factorization for exclusive processes leads to the
amplitude of this process been proportional to $G_A(x,Q^2)$
\cite{FMS93}. For small $x$ and $t$
 \begin{equation}
 {d\sigma^{\gamma^*_L+A \to V +A}\over dt}=F_A^2(t)
 {G_A^2(x,Q^2)\over G_N^2(x,Q^2)}{d\sigma^{\gamma^*_L+N\to V+N}\over
 dt}.
 \end{equation}
So we expect that the color transparency regime  for $x\ge 0.02$
(with a small enhancement at $x\sim 0.1$ due to enhancement of
 $G_A(x,Q^2)$ for these $x$)
would be followed by the color opacity regime for $x\le 0.01$.
As an illustration in Fig.9 we present the ratio of the cross section
of the $J/\psi$ and $\Upsilon$ production off nuclei with $A=12, 200$
and nucleon calculated under the assumption that the leading twist
 gives the dominant contribution in these processes.
 It is plotted as a
 function of x and normalized to the value of the ratio at $x=0.02$. In
 the calculation we use the analysis of \cite{FKS98} which indicates
 that $Q_{eff}^2 \approx 5 (40) GeV^2$ for $J/\psi$($\Upsilon$)
 production.  One can see from the figure that we expect onset of the
 Color Opacity regime for $J/\psi$ starting at $x\sim 0.01$. The Color
 Opacity effect remains quite significant for production of $\Upsilon$.

\section{Nuclear effects in the 
inclusive leading hadron spectra in eA collisions}

Numerous data on hadron-nucleus scattering at fixed target
energies indicate that the multiplcities of the leading hadrons
$N_A(z)\equiv {1 \over \sigma_{tot}(aA)}{d\sigma(z)^{a+A\to h +X}\over dz}$
decrease with increase of A. Here $z$ is the light-cone fraction of the 
projectile $"a" $ momentum carried by the hadron ``h''.
On the contrary,  the QCD factorization theorem for
the inclusive hadron production in DIS 
implies
that in the case of electron-nucleus scattering no such dependence 
should be present in the  DIS case. 
This indicates that there should be a interesting 
transition from the soft physics dominating in the interactions of
 real photons with nuclei  to the hard physics
in the inclusive hadron production in the DIS kinematics. It would be
 manifested in 
the disappearance
of the A-dependence of the leading spectra at large z:
\begin{equation}
N_A(z,Q^2)=N_N(z,Q^2), ~for~ z\ge 0.2, Q^2 \ge ~few ~GeV^2,  
\end{equation}
At small x a new interesting phenomenon should emerge due to
 the presence of diffraction and nuclear shadowing for smaller $z$.
Indeed, the diffraction originates from the presence 
in the wave function of $\gamma^*$ of partons with relatively
small
virtualities which screen the color of the leading parton(partons)
 with large  virtuality
and can rescatter elastically from  a target (several target nucleons in
the case of nuclear target). Inelastic interactions of these soft partons
with several nucleons should lead to a
plenty of new revealing phenomena in small $ x$ DIS
eA scattering, which  resemble hadron-nucleus scattering  but with a
shift in rapidity from $y_{max}(current)$
related to the average rapidities of these soft partons. This shift can be 
expressed through the average masses of the hadron states
produced in the diffraction:
\begin{equation}
y_{soft~partons} \sim  y_{max}- \ln (\left<M^2_{dif}\right>/\mu^2),
\end{equation}
where $\mu \sim 1 GeV$ is the soft  scale.
Partons with  these rapidities  will interact in multiple collisions and
loose their energy leading to a dip in the ratio 
$\eta_A(y)\equiv {N_A(y)/N_p(y)}$. At the same time these multiple 
interactions should generate a larger multiplcities at smaller rapidities.
Application of the AGK rules indicates that 
for $y \le y_{soft~partons} -\Delta$, where $\Delta = 2-3$
the hadron multiplicity in the case of nuclei will be enhanced
by the factor:
\begin{equation}
\eta_A(y)={AF_{2p}(x,Q^2) \over F_{2A}(x,Q^2) }.
\end{equation}
At the rapidities close to the nuclear rapidities a further increase of 
$\eta_A(y)$ is possible due to formation of hadrons inside the nucleus.
A sketch of the expected rapidity dependence of $\eta_A(y)$ is presented 
in Fig.10.

One also expects a number of phenomena due to long range correlations 
in rapidity. This includes: (a)  Local fluctuations of multiplicity
in the central rapidity region, e.g. the observation of a broader 
distribution of the number of particles per unit rapidity,  due to 
fluctuations of the number of wounded nucleons \cite{FSAGK}. 

These fluctuations should be larger for the hard processes induced by gluons,
for example the direct photon production of two high $p_t$ dijets.
(b) Correlation of the central multiplicity with the multiplicity of 
neutrons in the forward
neutron detector, etc.

Another important manifestation of nuclear shadowing is a
 large probability  of diffractive final states for small x.
 In the case of the
generic $e+A$ scattering this probability would reach $\sim 35\%$ for
$ A\sim 200$ \cite{FSAGK}. 
Due to a larger effective cross section 
of interaction in processes induced by the hard 
interactions with gluons,
this effect should be even more pronounced in the lepton scattering
processes where  a
hard process corresponds to hard $\gamma^*-g$ 
interaction. Hence for example in the charm electroproduction
of a heavy nucleus 
we expect about a half of the events to originate from the
 coherent diffraction where the nucleus remains intact \cite{FS981}.
Also one expects
more strong filtering out of the gluon dominated diffraction than  the
diffraction dominated by the coupling to the quarks.

\section{Comparison with other approaches}
Several  approaches were developed over last decade to the
dynamical calculation of  
nuclear shadowing phenomenon in DIS \footnote{A phenomenological approach 
based on fitting the existing nuclear data and imposing the momentum 
and baryon sum rules in the spirit of \cite{FLS} was pursued in
 \cite{Eskola93,eskola}.}

First group of approaches is based on the 
Gribov work \cite{gribov} which established
connection between the diffraction
and nuclear shadowing. Among these considerations
\cite{FS88,kwi,FS89,FLS,NZ,Piller,Kop,FSAGK,FSAGK,Orsay}
 the one of \cite{Orsay}
is the most detailed and 
comes closest to our analysis in the case of $F_{2A}$. The analysis is
based on
  the fit 
to the HERA diffractive data on the $e+p\to e+X+p$
reaction  within  the model developed by the authors
and a set of assumptions about higher order screening effects which
 are anyway rather small for $F_{2A}$. They demonstrate that 
the model can well describe the NMC data at 
$x \sim 0.01$ and give predictions for the HERA kinematics. The main 
differences from our approach are  the use of the model
for diffraction 
 which does not explicitly satisfy the QCD factorization theorem
for diffraction in DIS
and neglect by the  effects of enhancement of gluon
distributions in nuclei at $x\sim 0.1$.
Besides  the gluons appear to play a rather small role in their model
 of $ep$ diffraction leading to the expectation of 
the gluon shadowing smaller than in the case of $F_{2A}$ as compared
to the larger shadowing for gluons expected in our analysis.

In the case of hadron-nucleus scattering
both total cross section and inelastic diffraction can be described 
based on the idea of the fluctuations of the interaction strength 
in the projectile treating interaction of each component in the
eikonal approximation, for the review and references see \cite{FMSrev}.
There are a several models where a similar approach has been  applied to the
calculation of the 
nuclear shadowing by introducing the impact parameter $q \bar q$
virtual photon wave function $\Psi_{\gamma}(b)$
and introducing the cross section of  
the $q \bar q$-N interaction for the  fixed $b$, see e.g. \cite{RTV,KM}
and references therein. 
So far it was assumed in these models that
$\sigma_{q\bar q -N}=cb^2$. hence the QCD evolution which leads to a 
fast increase of $\sigma_{q\bar q -N}$ with incident energy was neglected.
Moreover in this approximation shadowing for the small b configurations 
is a higher twist effect, leading to an expectation of lack of the 
leading twist shadowing for
$\sigma_L$ and lack of gluon shadowing.

Another group of approaches uses the infinite momentum picture and 
treats all the process within the perturbative QCD. 
To avoid problems with positivity of the cross section
for $A_{eff}/A\le 0.75$ (see discussion in section XX) one has 
to include
interactions with $N\ge 3$ nucleons. 
The current models which include such interactions 
 (see \cite{AGL,HLS,JX} and references therein) assume 
that all shadowing is generated perturbatively and do not include
information about diffractive processes. Qualitative expectations 
of these models are a rather large gluon shadowing and significant 
nonlinear effects
in the evolution of the parton densities.
It would be interesting to compare two approaches after 
the leading twist shadowing effects are implemented and constraints 
following from the HERA diffractive data are taken into account.

\section{Conclusions}
 We have demonstrated that dominance of gluons in
the 
"Pomeron"
diffractive parton densities leads to a large enhancement
of the nuclear gluon shadowing in a wide range of $x,Q$.
Gluon shadowing of this magnitude will strongly affect the first stage
of the heavy ion collisions at LHC, lead to a number of
Color Opacity phenomena in the HERA kinematics for
the $eA$ collisions.
Study of the gluon shadowing for heavy nuclei may allow to enhance
contribution of the small interaction strengths, allowing to
unmask PQCD physics in the $eA$ collisions at HERA.
The  studies of coherent diffraction off nuclei
and hadron production in the inclusive eA scattering will
 provide complementary handles for studying the small x dynamics.

\acknowledgments{One of us (M.S.) would like to thank DESY
for the hospitality during the time this work was done.
We thank J.C.Collins and J.Whitmore for discussion of the
diffractive production.
We are indebted to A.Freund and V.Guzey for help
in running CTEQ QCD evolution code and to M.Zhalov for help with
figures. This work is supported in part by the U.S. Department of
Energy and BSF.}

\newpage

\vspace{1.0cm}
\begin{figure}
\centerline{
\epsfig{file=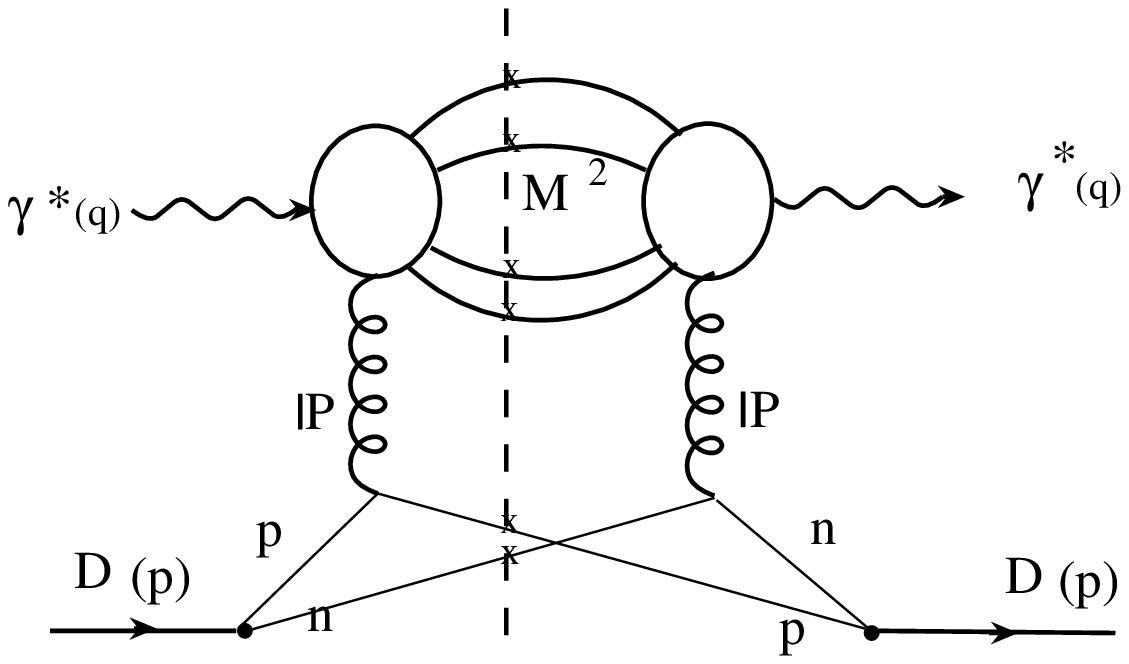,width=10.0cm,height=10.0cm}}
\caption{Double scattering diagram
for the $\gamma^*D$ scattering corresponding to a diffractive final state}
\end{figure}

\newpage

\vspace{1.0cm}
\begin{figure}
\centerline{
\epsfig{file=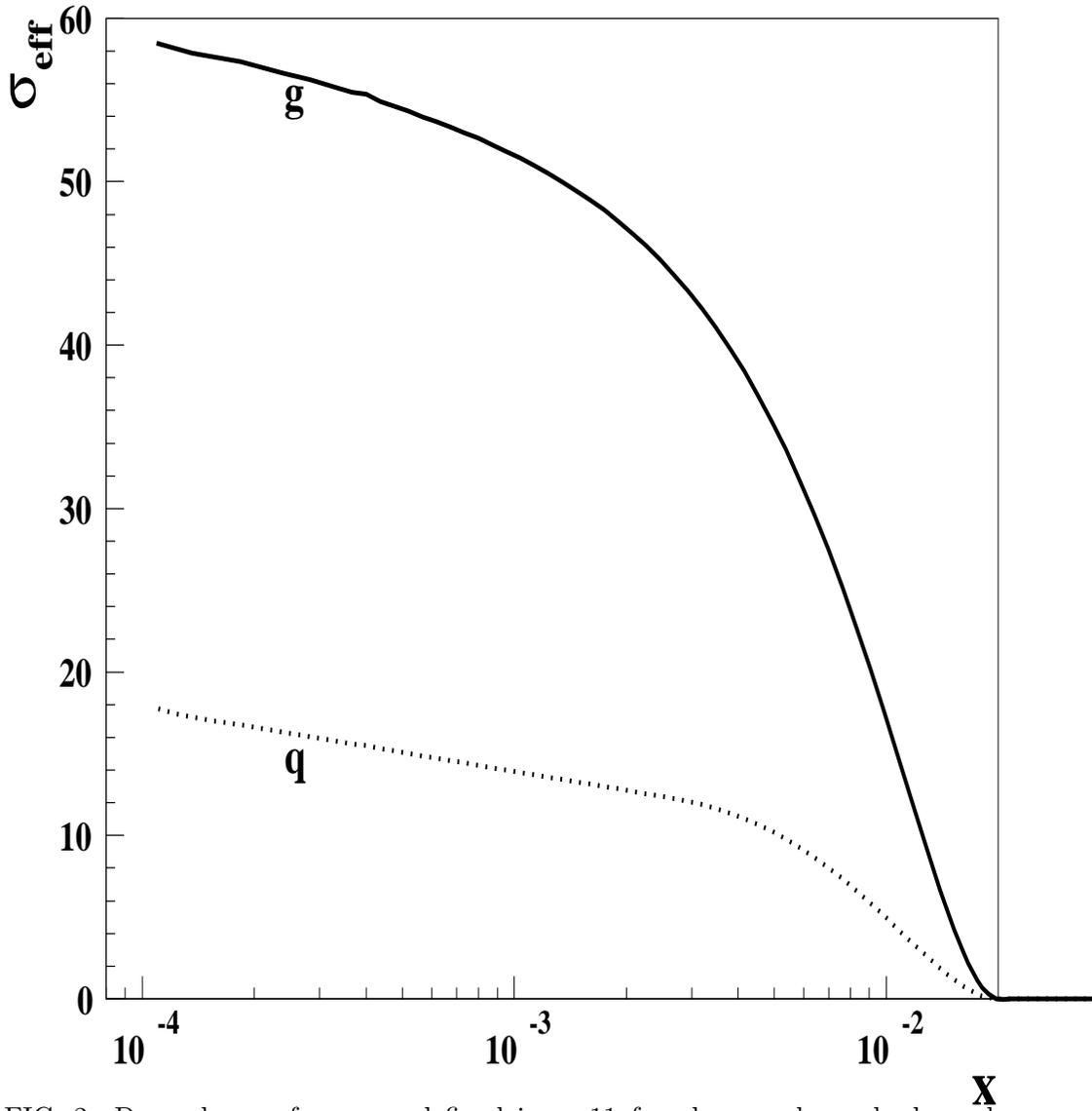,width=15.0cm,height=15.0cm}}
\caption{Dependence of $\sigma_{eff}$ as defined in eq.\ref{sidat}
for gluon and quark  channels
on $x$ for  $Q=Q_0=$ 2 GeV.}
\end{figure}

\newpage

\vspace{1.0cm}

\begin{figure}
\centerline{
\epsfig{file=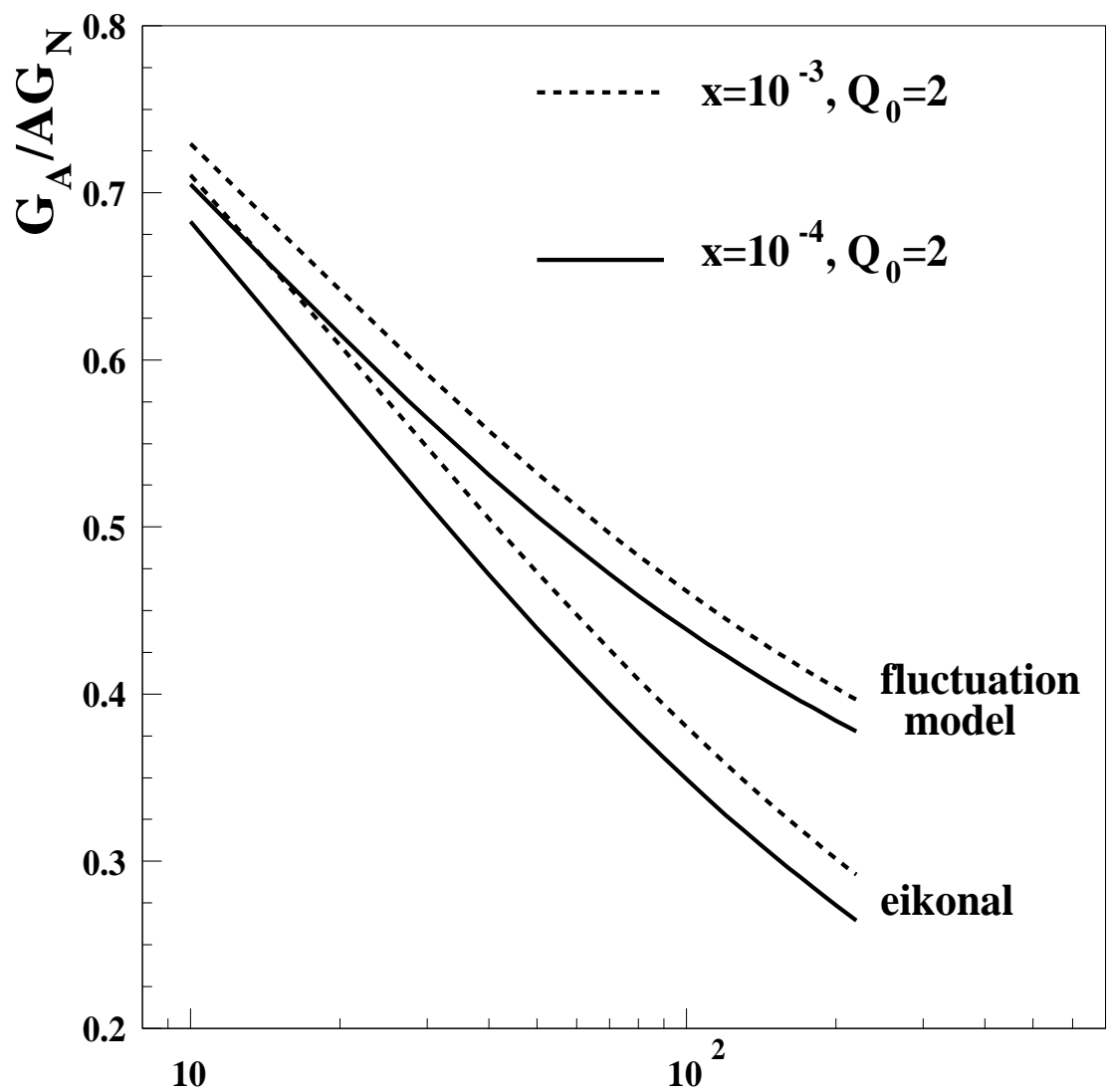,width=15.0cm,height=15.0cm}
}

\caption{Comparison of the gluon shadowing calculated in
the quasieikonal model and the fluctuation model
for $x=10^{-3},10^{-4}$.}
\end{figure}
\newpage

\vspace{1.0cm}

\begin{figure}
\centerline{
\epsfig{file=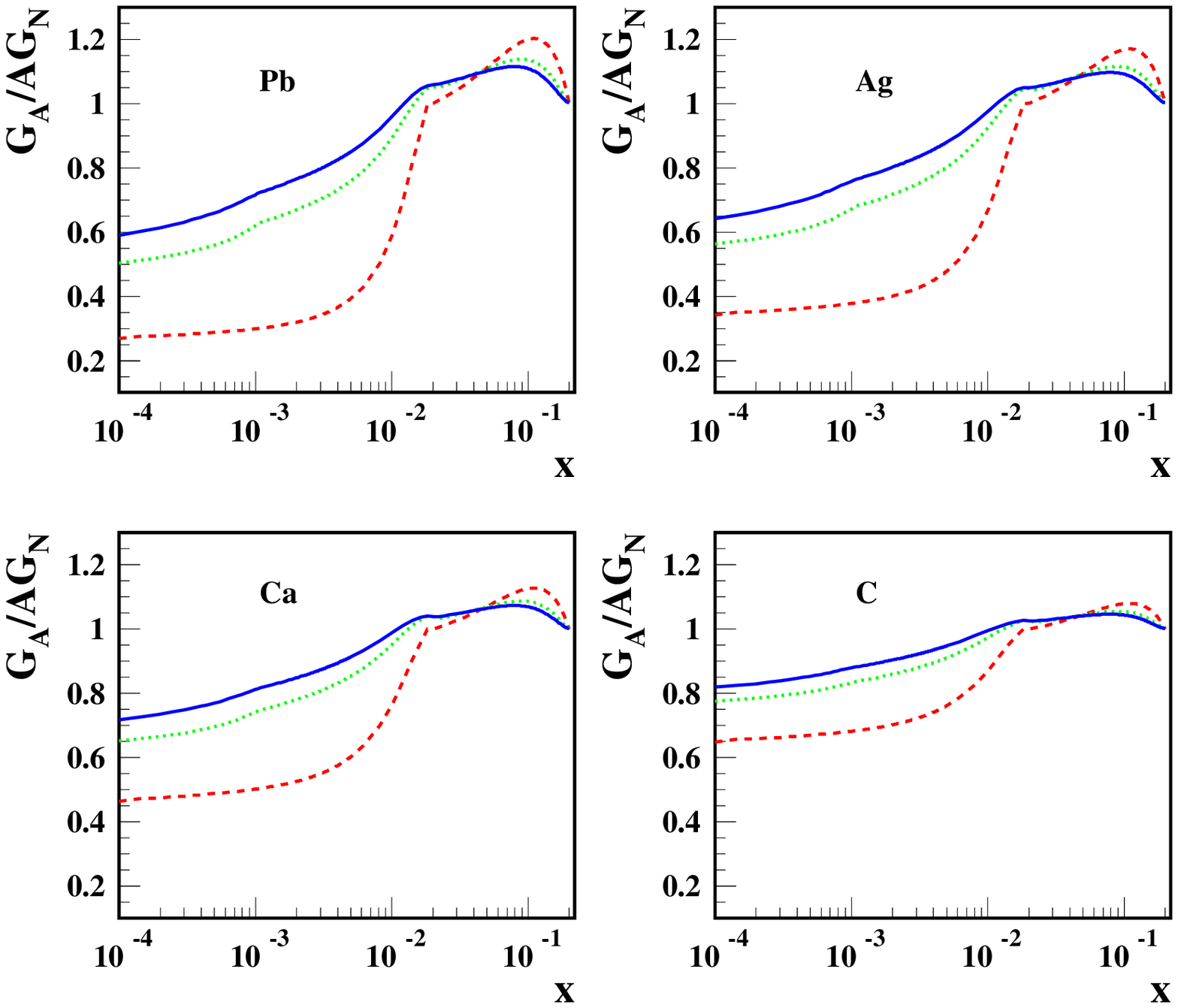,width=15.0cm,height=15.0cm}
}
\caption{Dependence of $G_A/AG_N$ on $x$
for Q=2,5,10 GeV  (dashed, dotted, solid curves) calculated in
the quasieikonal model.}
\end{figure}
\newpage

\vspace{1.0cm}

\begin{figure}
\centerline{
\epsfig{file=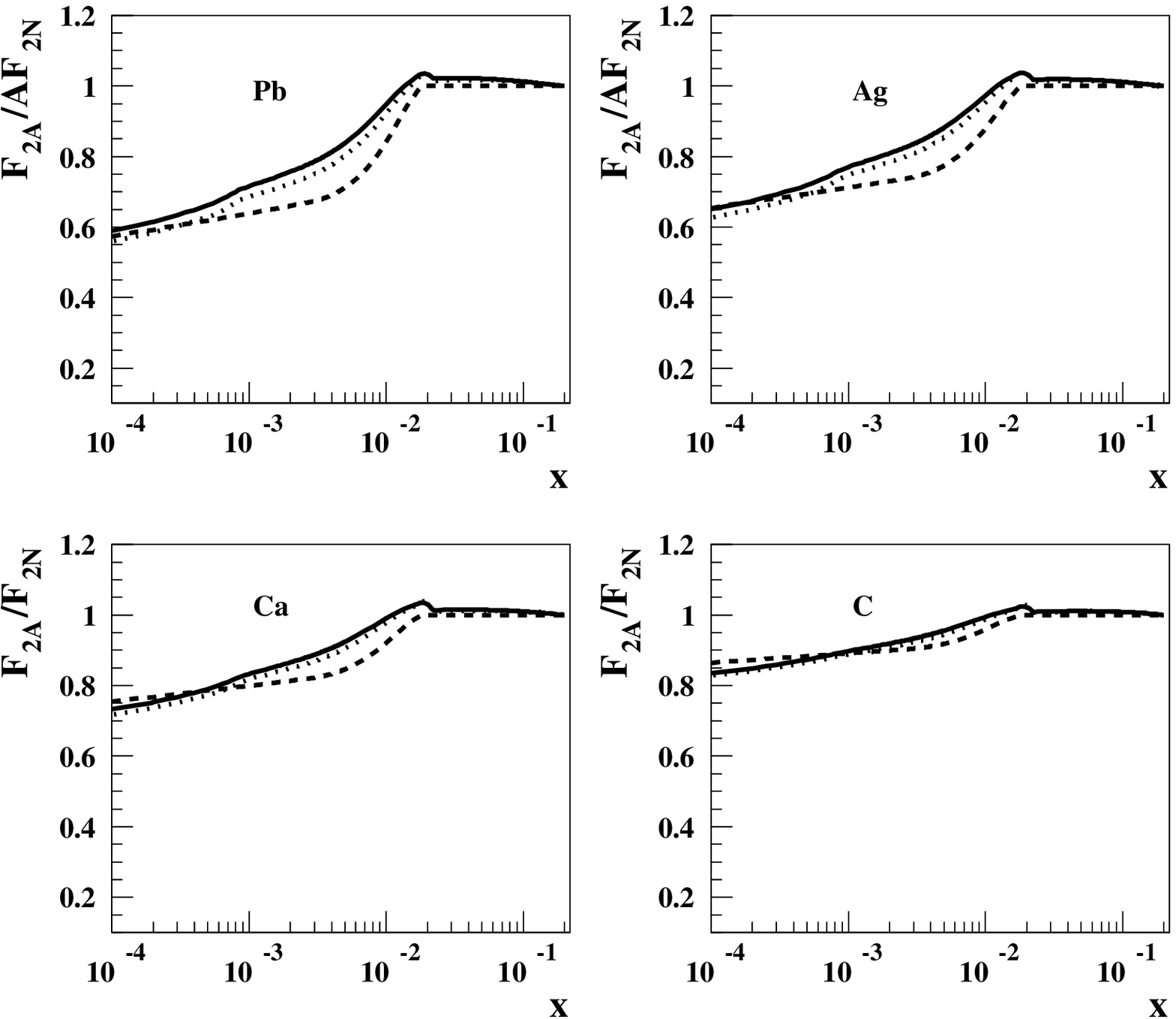,width=15.0cm,height=15.0cm}
}
\caption{Dependence of $F_{2A}/AF_{2N}$ on $x$ for Q=2,5,10 GeV
 (dashed, dotted and solid curves)
calculated in the quasieikonal model.}
\end{figure}
\newpage

\vspace{1.0cm}

\begin{figure}
\centerline{
\epsfig{file=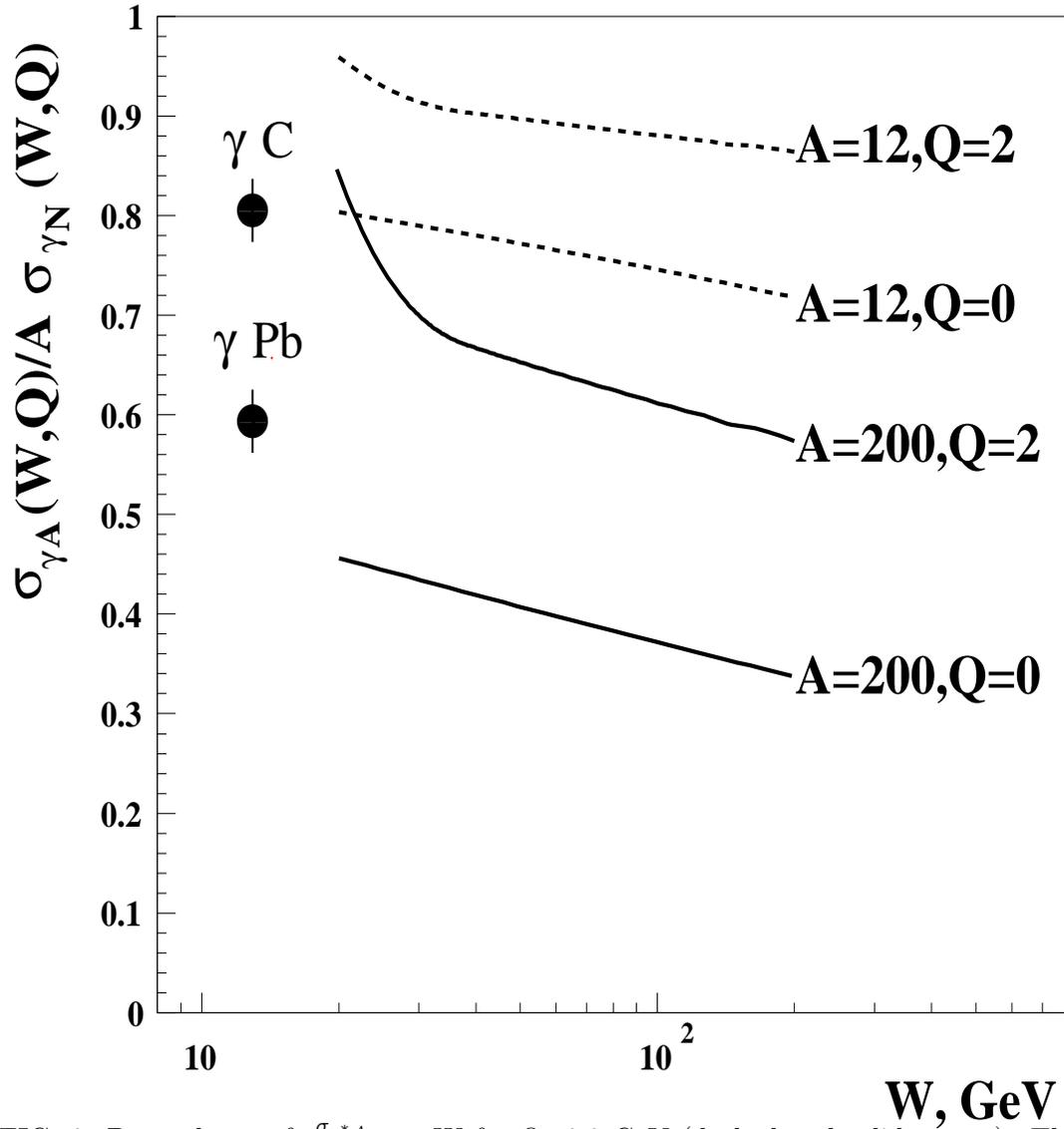,width=15.0cm,height=15.0cm}
}
\caption{Dependence of 
${\sigma_{\gamma^* A}\over A \sigma_{\gamma^* N}}$
on $W$ for Q=0,2 GeV (dashed and solid curves). 
 The data points are from \protect\cite{Caldwell}.
They are corrected for the small effect of the nuclear shadowing 
in the deuteron.}
\end{figure}

\newpage

\vspace{1.0cm}

\begin{figure}
\centerline{
\epsfig{file=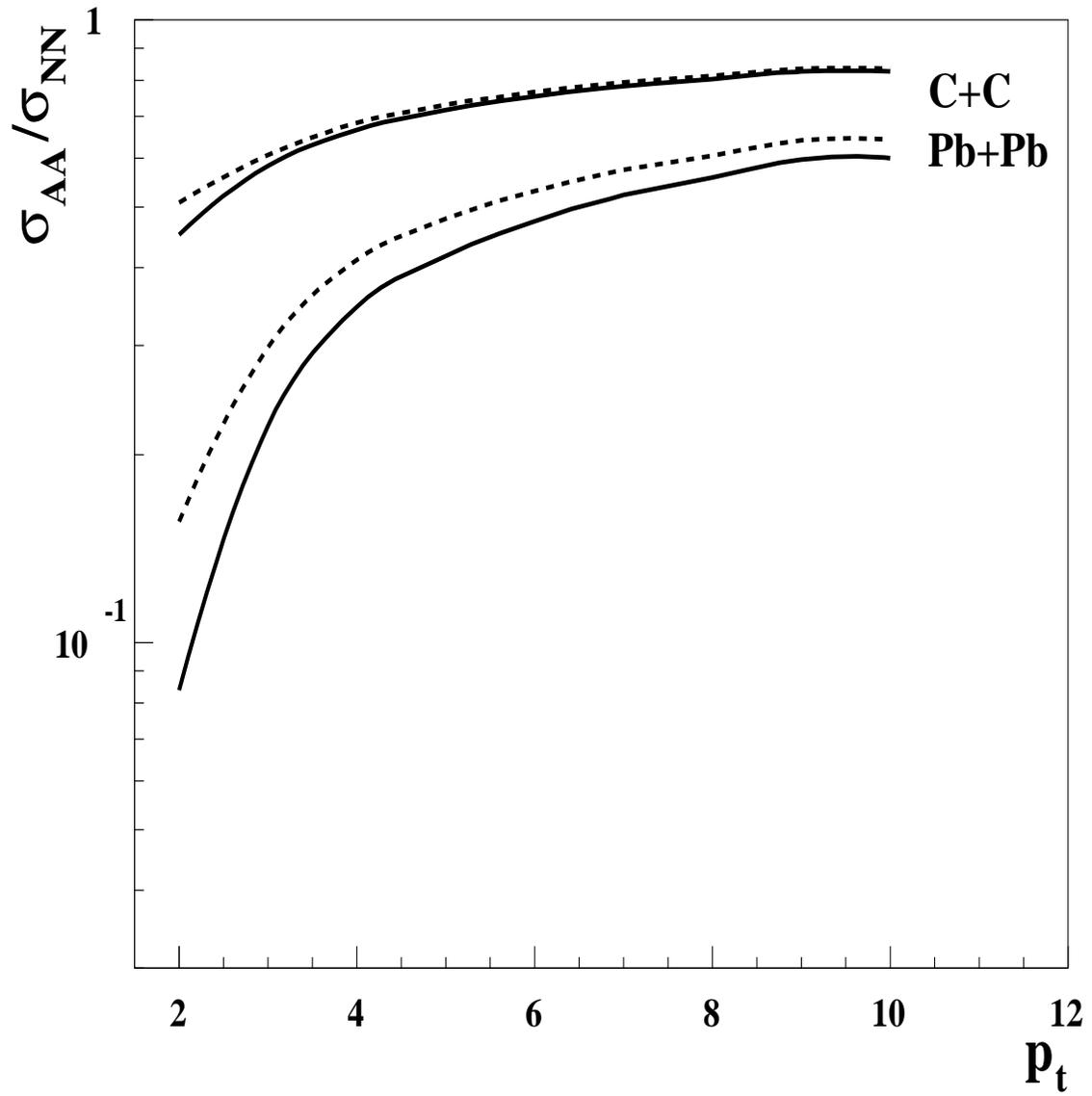,width=15cm,height=15cm}}

\vspace{1.0cm}

\caption{Suppression of the jet production in $AA$ collisions
due to gluon shadowing
at $y=0$ calculated in the quasieikonal and fluctuation models
(solid and dashed curves).}
\end{figure}
\newpage

\vspace{1.0cm}

\begin{figure}
\centerline{
\epsfig{file=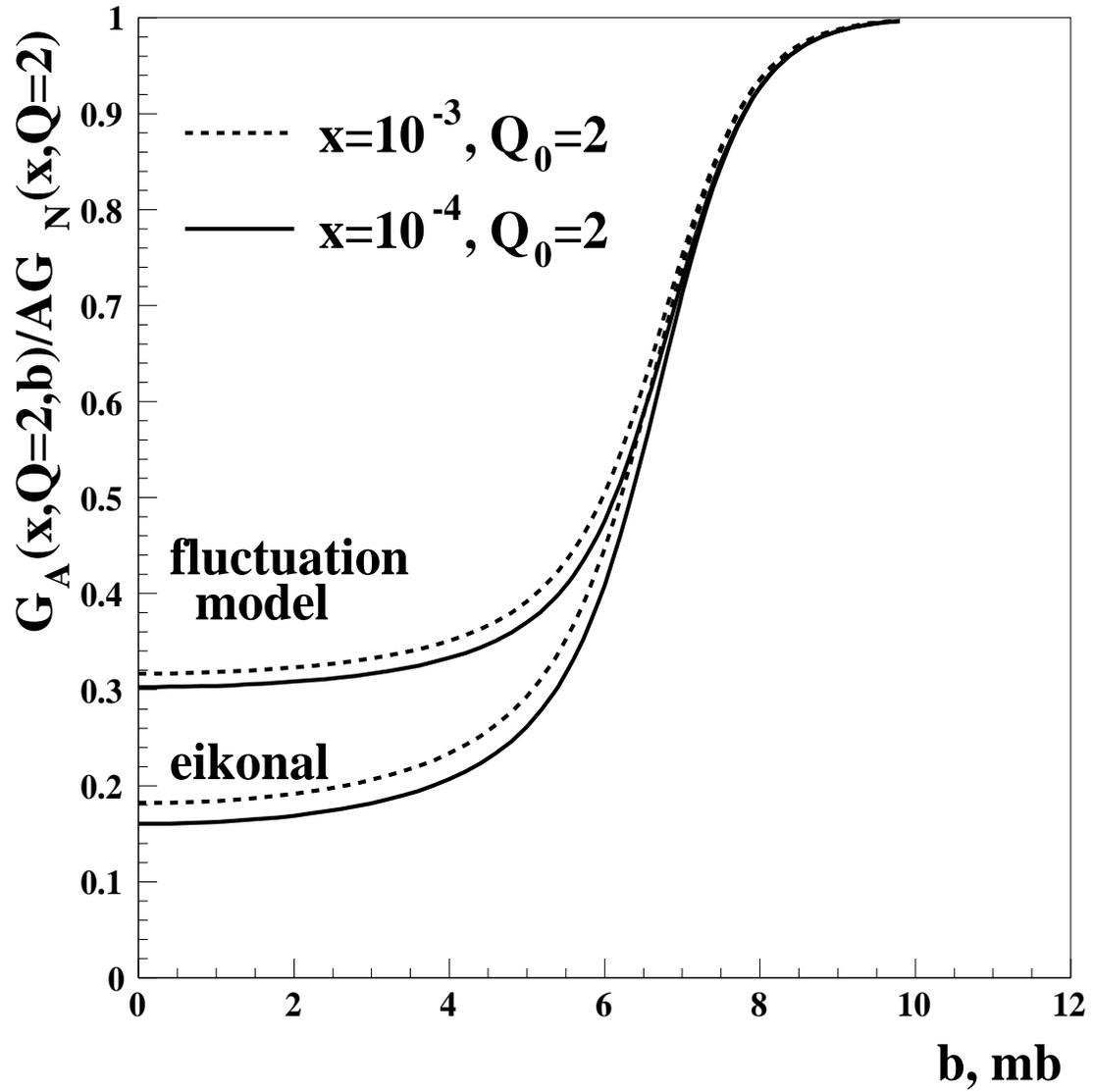,width=15cm,height=15cm}}

\vspace{1.0cm}

\caption{Impact parameter dependence of the gluon shadowing
for the scattering of $Pb$ for $Q=2$GeV and $x=10^{-3},10^{-4}$
calculated in the quasieikonal and fluctuation models.}
\end{figure}
\newpage

\vspace{1.0cm}

\begin{figure}
\centerline{
\epsfig{file=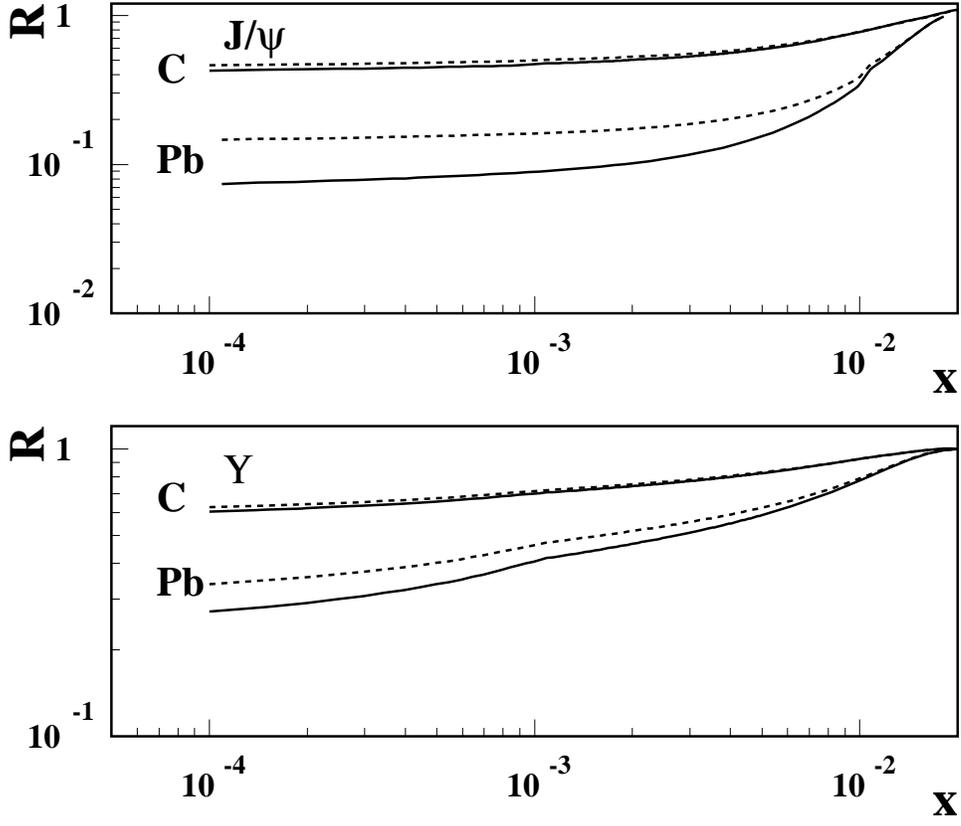,width=15cm,height=15cm}}

\vspace{1.0cm}

\caption{Color opacity effect for the ratio of the coherent production
of $J/\psi$ and $\Upsilon$ from carbon(lead)
and a nucleon normalized to the value of this ratio at $x=0.02$
 calculated in the quasieikonal and fluctuation models
(solid and dashed curves).}
\end{figure}
\newpage

\begin{figure} 
    \begin{center} 
        \leavevmode 
        \epsfxsize=1.00\hsize 
        \epsfbox{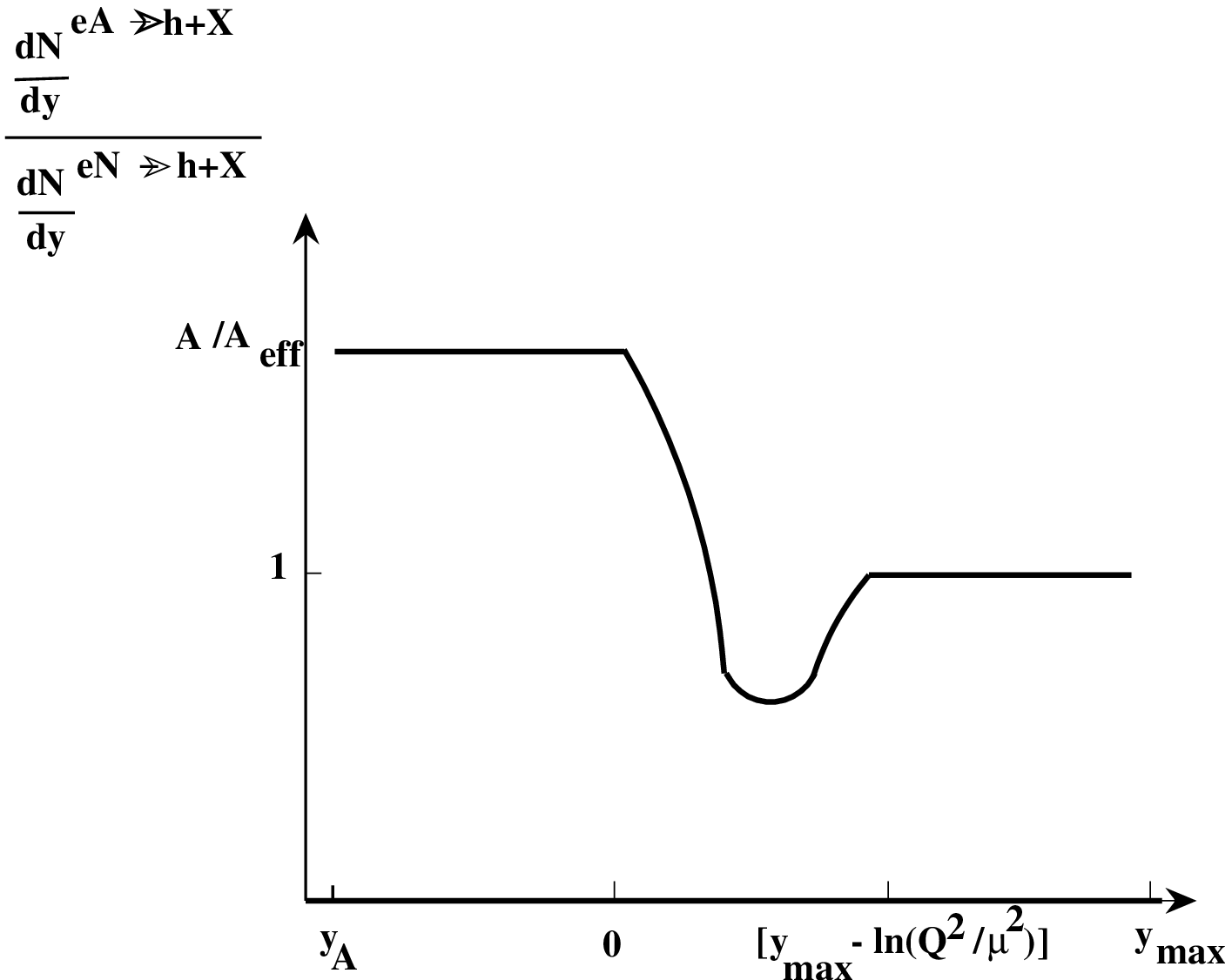} 
    \end{center} 
\caption{Sketch of the A dependence of the hadron multiplicity 
 ${dN \over dy}$ for small x
plotted as a function of hadron rapidity.}
\end{figure} 

\vspace{-0.4cm}
\noindent


\begin{thebibliography}{99}
\bibitem{gribov} V.N.Gribov, Sov.J.Nucl.Phys. {\bf 9} 1969) 369;
Sov.Phys.JETP {\bf 29} 1969, 483; ibid {\bf 30} (1970) 709.
\bibitem{FS88} L. L. Frankfurt and M. Strikman, Phys.Rep. {\bf 160}
(1988) 235.
\bibitem{kwi}J.Kwiecinski and B.Badelek, Phys.Lett. {\bf 208} (1988)508;
J.Kwiecinski, Z.Phys.{\bf C45} (1990) 461.
\bibitem{FS89}  L. L. Frankfurt and M.I.Strikman,
Nucl. Phys. {\bf B316} (1989)340.
\bibitem{FLS} L. L. Frankfurt, M. Strikman and S. Liuti, Phys. Rev.
 Lett. {\bf 65} (1990) 1725.

\bibitem{NZ} N. N. Nikolaev and B. G. Zakharov, Z. Phys. {\bf C49}
(1991) 607.

\bibitem{Piller}G. Piller, G. Niesler, W. Weise, Z. Phys. {\bf A358}
 (1997) 407.
\bibitem{Kop}B. Kopeliovich, B.Povh,
 Phys.Lett.
{\bf  B367}(1996) 329.
\bibitem{FSAGK} L.Frankfurt and M.Strikman, Phys.Lett. {\bf B382} (1996) 6.
\bibitem{Barone}V. Barone, M. Genovese,
Phys. Lett.{\bf  B412}
(1997) 143.

\bibitem{Orsay}A. Capella, A. Kaidalov, C. Merino,
D. Pertermann, J. Tran Thanh Van, Eur. Phys. J.{\bf C5}  (1998) 111.
\bibitem{Eskola93} K.Eskola, Nucl.Phys.{\bf B400}(1993)240.
\bibitem{FS981}L.Alvero,  L. L. Frankfurt and M. Strikman,
hep-ph@xxx.lanl.gov - 9810331, Europ.J.Phys.{\bf A 5}(1999) 97.
\bibitem{FLS93} L. L. Frankfurt, M. Strikman and S. Liuti,
Proceedings of XIII Particle and Nuclear Physics Conference, Perugia, Italy (1993) p. 342-343.
\bibitem{eskola}K.J. Eskola, V.J. Kolhinen, P.V. Ruuskanen,
Nucl. Phys. B535 (1998) 351-371.

\bibitem{Collins}J. C.Collins,Phys.Rev.{\bf D57} (1998) 3051.
\bibitem{AFS} H. Abramowicz, L.L. Frankfurt,
 and M. Strikman DESY-95-047,
 March 1995; Proceedings of SLAC Summer Inst., 1994, pp. 539-574.

\bibitem{ACW}L. Alvero, J. C. Collins, J.J. Whitmore,
hep-ph@xxx.lanl.gov - 9806340,  hep-ph@xxx.lanl.gov - 9805268.
\bibitem{MV}L.McLerran and R.Venugopalan, Phys. Rev. {\bf D49}(1994) 2233;
{\bf D49}(1994) 3352;{\bf D50}(1994)2225.
\bibitem{Mueller99}A.H. Mueller, hep-ph@xxx.lanl.gov - 9902302.
\bibitem{IS}G. Ingelman and P. E.Schlein,
Phys.Lett. {\bf 152B} (1985) 256.
\bibitem{AGK} V.Abramovski$\breve{{\rm i}}$, V. N. Gribov, and O. V.
 Kancheli, Sov. J. Nucl. Phys. {\bf 18}, (1974) 308.

\bibitem{VDM}T.H. Bauer, R.D. Spital, D.R. Yennie, F.M. Pipkin,
 Rev.Mod.Phys.{\bf 50}
(1978) 261, ERRATUM-ibid. {\bf 51} (1979) 407.


\bibitem{FMS93}L. Frankfurt, G.A. Miller and M. Strikman, Phys. Lett.
 {\bf B304}~(1993)~1;
S. J. Brodsky, L. Frankfurt, J.F. Gunion,
 A.H. Mueller, M. Strikman, Phys.Rev. {\bf D50}
 (1994) 3134-3144.
\bibitem{FKSpi}L. Frankfurt, W.Koepf, M.Strikman, Phys.Lett. {\bf B405}
(1997) 367.
\bibitem{Dino}D.Gualianos, hep-ph 9806363.
\bibitem{ZEUSdiff}ZEUS Collaboration (J. Breitweg, et al.), DESY-98-084,
Eur. Phys. J. {\bf C6} (1999) 43.
\bibitem{FP} E. Feinberg and Ya. Pomeranchuk, Suppl. Nuovo Cimento {\bf 111} (1956) 652.
\bibitem{GW} M. Good and W. Walker, Phys. Rev. {\bf D120} (1960) 1857


\bibitem{slopezeus}ZEUS collaboration, Contributed paper 972,
 XXIX International Conference on High Energy Physics,
 Vancouver, 23-29 July 1998.



\bibitem{MQ}A.H. Mueller and J.W.Qiu, Nucl.Phys.{\bf B268} (1986)427.
\bibitem{FKS96}L. Frankfurt, W.Koepf, M.Strikman, Phys.Rev.{\bf D54}
(1996) 3194.


\bibitem{EQW} K.J. Eskola, Jian-wei Qiu, Xin-Nian Wang,
Phys.Rev.Lett.{\bf 72} (1994)36.
\bibitem{FGS}L.Frankfurt, V.Guzey, M.Strikman, Phys.Rev.
 {\bf D58} (1998)94039.


\bibitem{GRV}M. Gluck, E. Reya, A. Vogt, Z.Phys. {\bf C53} (1992)127.
\bibitem{Pirner}T. Gousset and H.J. Pirner, Phys.Lett. {\bf B375} (1996)349.
\bibitem{Chapin}T.J.Chapin et al, Phys.Rev.D31,17(1985).
\bibitem{H1}H1, S.Aid et al., Z. Phys. {\bf C69} (1995) 27.
\bibitem{Caldwell}D.Caldwell et al, Phys.Rev.Lett. {\bf 42} 91979) 553.
\bibitem{FLSbnl}L.L. Frankfurt, M.I. Strikman and S. Liuti,
 in Proceedings of 4 BNL Workshop on Relativistic Heavy Ion Collisions,
July 1990, BNL 52262, p. 103-118.

\bibitem{FKS98}L. Frankfurt, W.Koepf, M.Strikman, Phys.Rev.{\bf D57}
(1998) 512-526.


\bibitem{FMSrev} L. L. Frankfurt, G. A. Miller and M. Strikman, Ann. Rev.
of Nucl.
and Particle Phys.~44~(1994)~501.
\bibitem{RTV}B.Z. Kopeliovich, J. Raufeisen
A.V. Tarasov,  Phys.Lett.{\bf B440} (1998)151; 
J. Raufeisen, A.V. Tarasov, O.O. Voskresenskaya, 
e-Print Archive: hep-ph/9812398 
\bibitem{KM}Yu. V. Kovchegov, L. McLerran
e-Print Archive: hep-ph/9903246.
\bibitem{AGL}A.L. Ayala, M.B. Gay Ducati,
E.M. Levin, Nucl.Phys.{\bf B493} (1997)305.
\bibitem{HLS} Zheng Huang, Hung Jung Lu, Ina Sarcevic,
 Nucl.Phys.{\bf A637} (1998) 79.
\bibitem{JX} J. Jalilian-Marian, Xin-Nian Wang
e-Print Archive: hep-ph/9902411. 

\end{thebibliography}
\end{document}